\documentclass[aps,prd,twocolumn,superscriptaddress,showpacs,floatfix]{revtex4}
\usepackage{bm}
\usepackage[dvips]{graphicx}

\usepackage{amsmath,amssymb,times}

\newcommand{\bequ}{\begin{equation}}
\newcommand{\eequ}{\end{equation}}
\newcommand{\bea}{\begin{eqnarray}}
\newcommand{\eea}{\end{eqnarray}}

\DeclareSymbolFont{boldletters}{OML}{cmm} {b}{it}
\DeclareSymbolFontAlphabet{\mathbit}{boldletters}
\DeclareMathSymbol{\alpha}{\mathalpha}{letters}{"0B}
\DeclareMathSymbol{\beta}{\mathalpha}{letters}{"0C}
\DeclareMathSymbol{\gamma}{\mathalpha}{letters}{"0D}
\DeclareMathSymbol{\delta}{\mathalpha}{letters}{"0E}
\DeclareMathSymbol{\epsilon}{\mathalpha}{letters}{"0F}
\DeclareMathSymbol{\zeta}{\mathalpha}{letters}{"10}
\DeclareMathSymbol{\eta}{\mathalpha}{letters}{"11}
\DeclareMathSymbol{\theta}{\mathalpha}{letters}{"12}
\DeclareMathSymbol{\iota}{\mathalpha}{letters}{"13}
\DeclareMathSymbol{\kappa}{\mathalpha}{letters}{"14}
\DeclareMathSymbol{\lambda}{\mathalpha}{letters}{"15}
\DeclareMathSymbol{\mu}{\mathalpha}{letters}{"16}
\DeclareMathSymbol{\nu}{\mathalpha}{letters}{"17}
\DeclareMathSymbol{\xi}{\mathalpha}{letters}{"18}
\DeclareMathSymbol{\pi}{\mathalpha}{letters}{"19}
\DeclareMathSymbol{\rho}{\mathalpha}{letters}{"1A}
\DeclareMathSymbol{\sigma}{\mathalpha}{letters}{"1B}
\DeclareMathSymbol{\tau}{\mathalpha}{letters}{"1C}
\DeclareMathSymbol{\upsilon}{\mathalpha}{letters}{"1D}
\DeclareMathSymbol{\phi}{\mathalpha}{letters}{"1E}
\DeclareMathSymbol{\chi}{\mathalpha}{letters}{"1F}
\DeclareMathSymbol{\psi}{\mathalpha}{letters}{"20}
\DeclareMathSymbol{\omega}{\mathalpha}{letters}{"21}
\DeclareMathSymbol{\varepsilon}{\mathalpha}{letters}{"22}
\DeclareMathSymbol{\vartheta}{\mathalpha}{letters}{"23}
\DeclareMathSymbol{\varpi}{\mathalpha}{letters}{"24}
\DeclareMathSymbol{\varrho}{\mathalpha}{letters}{"25}
\DeclareMathSymbol{\varsigma}{\mathalpha}{letters}{"26}
\DeclareMathSymbol{\varphi}{\mathalpha}{letters}{"27}
\DeclareMathSymbol{\Gamma}{\mathalpha}{letters}{"00}
\DeclareMathSymbol{\Delta}{\mathalpha}{letters}{"01}
\DeclareMathSymbol{\Theta}{\mathalpha}{letters}{"02}
\DeclareMathSymbol{\Lambda}{\mathalpha}{letters}{"03}
\DeclareMathSymbol{\Xi}{\mathalpha}{letters}{"04}
\DeclareMathSymbol{\Pi}{\mathalpha}{letters}{"05}
\DeclareMathSymbol{\Sigma}{\mathalpha}{letters}{"06}
\DeclareMathSymbol{\Upsilon}{\mathalpha}{letters}{"07}
\DeclareMathSymbol{\Phi}{\mathalpha}{letters}{"08}
\DeclareMathSymbol{\Psi}{\mathalpha}{letters}{"09}
\DeclareMathSymbol{\Omega}{\mathalpha}{letters}{"0A}

\begin{document}
\title{ Phase structure of two-color QCD at real and imaginary chemical 
potentials; \\ lattice simulations and model analyses}

\author{Takahiro Makiyama}
\email[]{12634019@edu.cc.saga-u.ac.jp}
\affiliation{Department of Physics, Saga University, Saga 840-8502, Japan}

\author{Yuji Sakai}
\email[]{ysakai@riken.jp}
\affiliation{Riken, Saitama 351-0198, Japan}

\author{Takuya Saito}
\email[]{tsaitou@kochi-u.ac.jp}
\affiliation{Integrated Information Center, Kochi University,
             Kochi 780-8520, Japan}

\author{Masahiro Ishii}
\email[]{ishii@email.phys.kyushu-u.ac.jp}
\affiliation{Department of Physics, Graduate School of Sciences, Kyushu University,
             Fukuoka 812-8581, Japan}
             
\author{Junichi Takahashi}
\email[]{takahashi@phys.kyushu-u.ac.jp}
\affiliation{Department of Physics, Graduate School of Sciences, Kyushu University,
             Fukuoka 812-8581, Japan}

\author{Kouji~~Kashiwa}
\email[]{kouji.kashiwa@yukawa.kyoto-u.ac.jp}
\affiliation{Yukawa Institute for Theoretical Physics, Kyoto University, Kyoto 606-8502, Japan}

\author{Hiroaki Kouno}
\email[]{kounoh@cc.saga-u.ac.jp}
\affiliation{Department of Physics, Saga University,
             Saga 840-8502, Japan}

\author{Atsushi Nakamura}
\email[]{nakamura@riise.hiroshima-u.ac.jp}
\affiliation{Research Institute for Information Science and Education, Hiroshima University,
             Higashi-Hiroshima 739-8527, Japan}

\author{Masanobu Yahiro}
\email[]{yahiro@phys.kyushu-u.ac.jp}
\affiliation{Department of Physics, Graduate School of Sciences, Kyushu University,
             Fukuoka 812-8581, Japan}

\date{\today}

\begin{abstract}
We investigate the phase structure of two-color QCD at both real and 
imaginary chemical potentials ($\mu$), 
performing lattice simulations and analyzing the data with 
the Polyakov-loop extended Nambu--Jona-Lasinio (PNJL) model.  
Lattice QCD simulations are done on an $8^{3}\times 4$ lattice 
with the clover-improved  two-flavor Wilson fermion action and 
the renormalization-group improved Iwasaki gauge action. 
We test the analytic continuation of physical quantities from 
imaginary $\mu$ to real $\mu$ by comparing lattice QCD results calculated 
at real $\mu$ with the result of analytic function the coefficients of which 
are determined from lattice QCD results at imaginary $\mu$. 
We also test the validity of the PNJL model by comparing 
model results with lattice QCD ones. 
The PNJL model is good in the deconfinement region, but less accurate 
in the transition and confinement regions. 
This problem is improved by introducing the baryon degree of freedom to 
the model. 
It is also found that the vector-type four-quark interaction is necessary 
to explain lattice data on the quark number density. 
\end{abstract}

\pacs{11.15.Ha, 12.38.Gc, 12.38.Mh, 25.75.Nq}
\maketitle
\section{Introduction}
\label{sec:intro}
Exploration of QCD phase diagram is one of the most important subjects in not only nuclear and particle physics but also cosmology and astrophysics.  
However, due to the complexity of fermion determinant, the first principle calculation, i.e., lattice QCD (LQCD) simulations are quite difficult at high quark 
number chemical potential $\mu$.  
The QCD partition function $Z$ at finite temperature $T$ and finite $\mu$ is expressed by 
\bea
Z=\int DU{\rm det}[M(\mu )]e^{-S_{\rm G}} ,
\label{eq_Z}
\eea
where $U_\mu~~~(\mu =1,2,3,4)$ and $S_{\rm G}$ are the link variables and the pure gauge action, respectively, and $M(\mu )$ is written 
as 
\bea
M(\mu )= \gamma_\mu D_\mu +m -\mu\gamma_4
\label{eq_Mmu}
\eea
with the covariant derivative $D_\mu$ and the quark mass $m$ in the continuum limit.  For later convenience, we regard $\mu$ as a complex variable. 
It is easy to verify 
\bea
\{{\rm det}[M(\mu )]\}^*={\rm det}[M(-\mu^*)]. 
\label{eq_sign}
\eea
Hence, the fermion determinant ${\rm det}[M(\mu )]$ is not real 
when $\mu$ is real, 
and the importance sampling technique does not work 
in the Monte Carlo simulations there. 
This is the well-known sign problem. 
Several methods were proposed so far to resolve this problem; 
these are the reweighting method~\cite{Fodor}, the Taylor expansion method~\cite{Allton,Ejiri_density}, the  analytic continuation from imaginary $\mu$ 
to real $\mu$~\cite{FP,D'Elia,D'Elia3,FP2010,Nagata,Takahashi,Chen}, 
the complex Langevin method~\cite{Aarts_CLE_1,Aarts_CLE_2,Sexty} 
and the  Lefschetz thimble theory~\cite{Aurora_thimbles,Fujii_thimbles}.   
However, these are still far from perfection. 

On the contrary, in two color QCD (QC$_2$D), the lattice simulations can be made at real and finite $\mu$, since the theory has no sign problem~\cite{Nakamura_nc2,Kogut_nc2,Muroya_nc2,Hands}.  
In fact, the following relation is obtained: 
\begin{align}
\det[M(\mu)]&=\det[(t_2 C \gamma_5)^{-1}M(\mu)(t_2 C \gamma_5)] 
\notag \\ 
&=(\det[M(\mu^*)])^* ,
\label{eq_sign_nc2}
\end{align}
where $t_2$ and $C=\gamma_2\gamma_4$ are the second Pauli matrix in color space and the charge conjugation matrix, respectively. 
Obviously, ${\rm det}[M(\mu )]$ is real when $\mu$ is real.   
Recently, Hands et al. analyzed the phase structure of QC$_2$D in a wide range of real $\mu$ by using two-flavor Wilson fermions~\cite{Hands_2,Cotter}. 
QC$_2$D can be also used to check the validity of methods proposed 
to resolve the sign problem. In fact, 
Cea et al.~\cite{Cea_1,Cea_2} tested the validity of analytic continuation 
from imaginary $\mu$ to real $\mu$ with staggered fermions.    

Equation~(\ref{eq_sign}) shows that ${\rm det}[M(\mu )]$ is real when $\mu$ is pure imaginary, i.e., $\mu =i\mu_{\rm I}=i\theta T$ for real variables $\mu_{\rm I}$ and $\theta$, so that LQCD simulations are feasible there.  
Observables at real $\mu$ are extracted from those at imaginary $\mu$ with analytic continuation. 
In the analytic continuation, we must pay attention to the structure of 
phase diagram in the imaginary $\mu$ region 
where QCD has two characteristic properties, 
the Roberge-Weiss (RW) periodicity and the RW transition~\cite{RW}.  
The QCD grand partition function has a periodicity of $2\pi/N_{c}$ in $\theta$:  
\bea
Z\left( \theta \right)=Z\left( \theta + \frac{2\pi k}{N_{c}} \right)
\eea
for integer $k$ and the number of color $N_{c}$.  
This periodicity was found by Roberge and Weiss and is then called the RW periodicity. 
Roberge and Weiss also showed that a first-order phase transition occurs 
at $T \ge T_{\rm RW}$ and $\theta =(2k+1)\pi/N_{c}$. 
This transition is named the RW transition, and $T_{\rm RW}$ is slightly 
larger than the pseudo-critical temperature $T_{c0}$ of deconfinement transition at zero $\mu$. 
These features are remnants of ${\mathrm Z}_{N_{c}}$ symmetry in the pure gauge limit. 
These properties are confirmed by LQCD simulations~\cite{FP,D'Elia,D'Elia3,FP2010,Nagata,Takahashi,Chen,Cea_1,Cea_2}. 

The RW periodicity does not mean that ${Z}_{N_{c}}$ symmetry is exact. 
Hence, there is no a priori reason that the order parameter for ${Z}_{N_{c}}$ symmetry such as the Polyakov loop $\Phi$ is zero in the confinement phase. 
In fact, in the case of $N_c =3$, the Polyakov loop is always finite 
even in the confinement phase, when $T$ is finite. 
However, the case of $N_c=2$ is special~\cite{Cea_1}.  
In this case, the action and the boundary conditions are invariant at $\mu_{\rm I}/T=(2k+1)\pi/2$ under the ${\cal C}Z_2$ transformation  composed of the ${\mathrm Z}_2$ transformation and charge conjugation~${\cal C}$~\cite{Kashiwa_nc2}. 
Due to this symmetry, the Polyakov loop becomes zero at low $T$ 
when $\mu_{\rm I}/T=(2k+1)\pi/2$. 
Paying attention to these characteristic features, Cea et al.~\cite{Cea_1,Cea_2} analyzed the validity of analytic continuation in QC$_2$D and found that 
lattice QC$_2$D (LQC$_2$D) data at real $\mu$ can be described by a suitable 
analytic function, when the coefficients of analytic function are 
determined from LQC$_2$D data at imaginary $\mu$.   

The results of LQCD at imaginary $\mu$ are also useful to determine the parameters of effective models, such as the Polyakov-loop extended Nambu--Jona-Lasinio (PNJL) model~\cite{Meisinger,Dumitru,Fukushima1,Ghos,Megias,Ratti1,Rossner}. 
Here we call this approach "imaginary chemical potential matching approach"~\cite{Kashiwa_meson}.  
It is known that the PNJL model can reproduce the results of LQCD at imaginary $\mu$, at least qualitatively, since the model has the RW periodicity and the RW transition~\cite{Sakai_IM,Kouno}. 
It was proposed~\cite{Sakai_vector} that the strength $G_{\rm v}$ of vector-type four-quark interaction~\cite{Kashiwa1,Sugano}, which is expected to be important for the physics of neutron stars,  may be determined from LQCD data at imaginary $\mu$; for the relation between neutron star properties and $G_{\rm v}$, 
see Ref.~\cite{Sasaki_NS} and references therein.  
In Refs.~\cite{Sakai_para} and~\cite{Kashiwa_nonlocal}, in fact, $G_{\rm v}$ is determined with this prescription.   
The validity of such a determination of parameters in effective models can be checked in QC$_2$D.           

In this paper, we study the phase structure of QC$_2$D at both 
real and imaginary $\mu$ by performing simulations 
on an $8^{3}\times 4$ lattice with the renormalization-group improved Iwasaki gauge action~\cite{Iwasaki,Iwasaki3} and the clover-improved  two-flavor Wilson fermion action~\cite{clover-Wilson} and analyzing the QC$_2$D data 
with the PNJL model.  
We first test the analytic continuation from imaginary $\mu$ to real $\mu$ 
by comparing LQC$_2$D data calculated at real $\mu$ with the result of 
analytic function the coefficients of which are determined from LQC$_2$D data 
at imaginary $\mu$. 
Such a test was tested in Refs. \cite{Cea_1,Cea_2} with staggered fermions. 
Here the test is made with clover-improved Wilson fermions 
by assuming a polynomial series in the deconfinement phase 
and a Fourier series in the confinement phase.

We second test the validity of the PNJL model by comparing 
LQC$_2$D results with model ones. 
The PNJL model is good in the deconfinement region, but less accurate 
in the confinement region. 
This problem is improved by introducing the baryon degree of freedom to 
the model. 
It is also found that the vector-type four-quark interaction is necessary 
to explain QC$_2$D data on the quark number density $n_{q}$.

This paper is organized as follows. 
Section~\ref{Lattice Simulation} presents the lattice action and the parameter setting used in our LQC$_{2}$D simulations. 
The definition of physical quantities is also presented.  
In Sec.~\ref{PNJLM}, the PNJL model is recapitulated.   
In Sec.~\ref{ACP}, numerical results of LQC$_{2}$D are shown and the analytical continuation of physical quantity from imaginary $\mu$ to real $\mu$ 
is tested. 
Comparison between LQC$_{2}$D data and PNJL results are made 
in Sec.~\ref{ComLP}. 
Section \ref{Summary} is devoted to a summary.

\section{Lattice Simulations}
\label{Lattice Simulation}

\subsection{Lattice action}

We use the renormalization-group-improved Iwasaki gauge action $S_{\rm G}$~\cite{Iwasaki,Iwasaki3} and the clover-improved two-flavor Wilson quark action $S_{\rm Q}$~\cite{clover-Wilson} defined by
\bea
S&=&S_{\rm G}+S_{\rm Q}, \\
S_{\rm G}&=&-\beta \sum_{x} \left(c_{0} \sum^{4}_{\mu<\nu;\mu,\nu=1}W^{1\times 1}_{\mu\nu}(x)\right. \nonumber \\
&&\left. +c_{1} \sum^{4}_{\mu\neq\nu;\mu,\nu=1}W^{1\times 2}_{\mu\nu}(x)\right), \\
S_{\rm Q}&=&\sum_{f=u,d}\sum_{x,y}\bar{q}^{f}_{x}M_{x,y}q^{f}_{y},
\eea
where $q$ is the quark field, $\beta=4/g^{2}$, $c_{1}=-0.331$, $c_{0}=1-8c_{1}$, and
\bea
M_{x,y}=&&\delta_{xy}-\kappa\sum^{3}_{i=1}\{(1-\gamma_{i})U_{x,i}\delta_{x+\hat{i},y}
\nonumber\\
&&+(1+\gamma_{i})U^{\dag}_{y,i}\delta_{x,y+\hat{i}}\} 
\nonumber \\
&&-\kappa\{e^{\mu}(1-\gamma_{4})U_{x,4}\delta_{x+\hat{4},y}
\nonumber\\
&&+e^{-\mu}(1+\gamma_{4})U^{\dag}_{y,4}\delta_{x,y+\hat{4}}\} 
\nonumber \\
&&-\delta_{xy}c_{\mathrm{sw}}\kappa\sum_{\mu<\nu}\sigma_{\mu\nu}F_{\mu\nu}.
\eea
Here $\kappa$ is the hopping parameter, $F_{\mu\nu}$ is the lattice field strength and $F_{\mu\nu}=(f_{\mu\nu}-f^{\dag}_{\mu\nu})/(8i)$ with $f_{\mu\nu}$ the standard clover-shaped combination of gauge links. 

The coefficient $c_{\mathrm{sw}}$ of clover term is determined by using a result obtained in a perturbative mean-field improved value $c_{\mathrm{sw}}=P^{-3/4}$~\cite{Lepage} with the plaquette $P$ calculated in one-loop perturbation theory, $P=1-0.3154\beta^{-1}$ for $N_c=2$~\cite{Iwasaki2}. 

\bigskip

\bigskip

\subsection{Parameter setting for simulations}

We denote temporal and spatial lattice sizes 
as $N_{t}$ and $N_{s}$, respectively. 
The Hybrid Monte-Carlo algorithm is used to generate full QC$_{2}$D configurations with two-flavor dynamical quarks. 
The simulations are performed on a lattice of $N_{s}^{3}\times N_{t} = 8^3 \times 4$.  
The step size of molecular dynamics is $\delta \tau = 0.02$ and 
the step number of the dynamics is $N_{\tau} = 50$. 
The acceptance ratio is more than 95\%. 
We generated 10,000 trajectories and removed the first 5,000 trajectories as thermalization for all the parameter set. 
The relation of parameters $\kappa$ and $\beta$ 
to the corresponding $T/T_{c0}$ 
is determined by finding the line of constant physics 
where the ratio of the pseudo-scalar (PS) meson mass $m_{\rm ps}$ 
to the vector meson mass $m_{\rm v}$ at $T=\mu =0$ are invariant; 
see Table \ref{table-para} for the relation.

\begin{table}[h]
\begin{center}
\begin{tabular}{|c|c|c|c|c|}
\hline 
$N_{s}$ & $N_{t}$ & $\beta$ & $\kappa$  & $T/T_{\mathrm{c0}}$
\\
\hline
~~$8$~~ & ~~$4$~~~ & ~~$0.60000$~~ & ~~$0.13782$~~ & 0.87783(585)~~~~\\ \cline{3-5}
             & ~~~~ & ~~$0.64000$~~ & ~~$0.13770$~~ & 0.94126(628)~~~~\\ \cline{3-5}
             & ~~~~ & ~~$0.66000$~~ & ~~$0.13751$~~ & 0.98577(657)~~~~\\ \cline{3-5}
             & ~~~~ & ~~$0.68000$~~ & ~~$0.13695$~~ & 1.02608(84)~~~~\\ \cline{3-5}
             & ~~~~ & ~~$0.70000$~~ & ~~$0.13677$~~ & 1.06282(709)~~~~\\ \cline{3-5}
             & ~~~~ & ~~$0.72000$~~ & ~~$0.13679$~~ & 1.11629(745)~~~~\\ \cline{3-5}
             & ~~~~ & ~~$0.74000$~~ & ~~$0.13567$~~ & 1.16526(77)~~~~\\ \cline{3-5}
             & ~~~~ & ~~$0.76000$~~ & ~~$0.13493$~~ & 1.20197(802)~~~~\\ \cline{3-5}
             & ~~~~ & ~~$0.78000$~~ & ~~$0.13443$~~ & 1.24298(829)~~~~\\ 
\hline 
\end{tabular}
\caption{
Summary of simulation parameters. $T_{\mathrm{c0}}$ is 
the pseudocritical temperature at $\mu=0$. 
In this parameter setting, the lattice spacing $a$ is about $1.38\sim 1.96$~{$\rm GeV^{-1}$} and the ratio $m_{\rm ps}/m_{\rm v}$ is 0.8 at $T=0$.}
\label{table-para}
\end{center}
\end{table}

\subsection{Physical observables}
\label{LSP}

In this paper, we calculate the Polyakov loop ($\Phi$),
 the quark number density ($n_q$) and the chiral condensate ($\sigma$). 
The quark number density is calculated by 
\bea
n_q={T\over{V}}{\partial\over{\partial \mu}}\log{Z}, 
\label{eq_nq}
\eea
where $V$ is the spatial volume, and $\sigma$ is by 
\bea
\sigma =\langle \bar{q}q\rangle ,  
\label{eq_sigma}
\eea
where $q$ is quark field and the $\langle O \rangle$ is the average value of physical quantity $O$. 
The chiral condensate suffers from the renormalization, and chiral symmetry 
is explicitly broken by Wilson fermions. This makes it difficult 
to deal with the absolute value of chiral condensate itself. 
We then consider a variation 
\begin{eqnarray}
\delta \sigma (T,\mu )=\sigma (T,\mu )-\sigma (T,0) .
\label{E_delta_sigma}
\end{eqnarray}

The Polyakov-loop operator is defined by 
\bea
L(\bm{x})={1\over{N_c}}\prod^{N_{t}}_{t=1}U_{4}(\bm{x},t)
\eea
with link variables $U_{\mu} \in \mathrm{SU}(2)$. 
The average value $\Phi$ of $L$ is related to the single static-quark free energy $F_q$ as 
\bea
 \Phi=\langle  L \rangle \sim e^{-F_q/T}  .
\eea
The Polyakov loop $\Phi$ is an order parameter of the confinement/deconfinement transition if $m$ is infinitely large. 
In fact, if $F_q$ is finite (infinite), $\Phi$ is finite (zero).  
The symmetry associated with the confinement/deconfinement transition 
is ${Z}_2$ symmetry under the transformation
\bea
U_4({\bf x},t )\to z_2(t)U_4({\bf x},t ), 
\label{eq_Z2}
\eea
where $z_2$ is the element of ${Z}_2$ group that depends only on the temporal coordinate $t$. 
Pure gauge action is invariant under this transformation while $L$ is not. 
Hence, $\Phi =\langle L \rangle$ is an order parameter of 
${Z}_2$ symmetry breaking.
Effects of dynamical quark break ${Z}_2$ symmetry explicitly and 
$\Phi$ is not a proper order parameter of 
confinement/deconfinement transition. 
As mentioned in the previous section, however, at $\theta =(2k+1)\pi/2$ 
the system is symmetric under the ${\cal C}Z_2$ 
transformation~\cite{Kashiwa_nc2}.  
Hence, $\Phi$ becomes an order parameter of the combined symmetry there.

\section{PNJL model}
\label{PNJLM}

Two-color QCD has Pauli-G\rm{\"{u}}rsey symmetry in the limit of $m=\mu=0$ \cite{pauli 1957,pauli 1958}. 
The PNJL Lagrangian of QC$_2$D is so constructed as to have the symmetry and 
is given by~\cite{Brauner,Kashiwa_nc2} 
\begin{eqnarray}
{\cal{L}}&=&\bar{q}(i \gamma^{\nu} D_{\nu}-m)q \nonumber \\
&+&G[(\bar{q} q)^2+(\bar{q} i \gamma_5 \vec{\tau} q)^2+|q^T C i \gamma_5 \tau_2 t_2 q|^2] \nonumber \\
&+&G_8[(\bar{q}q)^2+(\bar{q}i \gamma_5 \vec{\tau} q)^2+|q^T C i \gamma_5 \tau_2 t_2 q|^2]^2 \nonumber \\
&-&G_{\rm v}(\bar{q} \gamma^{\nu} q)^2-{\cal{U}}(\Phi),
\end{eqnarray} 
where $q$, $m$, $t_i$, $\tau_i$, $G$, $G_8$, $G_{\rm v}$ are 
the two-flavor quark fields, the current quark mass, the Pauli matrices in the color and the flavor spaces, the coupling constants 
of the scalar-type four-quark interaction, the scalar-type eight-quark 
interaction and the vector-type four-quark interaction, respectively.   
The potential ${\cal{U}}$ is a function of $\Phi$. 

The mean-field approximation leads us to 
the thermodynamical potential $\Omega$ as \cite{Brauner} 
\begin{eqnarray}
\Omega&=&-2 N_f \int \frac{d^3 p}{(2 \pi)^3} \sum_{\pm}[ E_{p}^++E_{p}^-+T(\ln{f^-}+\ln{f^+})] \nonumber \\
&+&U+{\cal{U}}(\Phi)
\end{eqnarray}
with
\begin{eqnarray}
f^{\pm}&=&1+2\Phi {\rm{e}}^{- \beta E_{p}^{\pm}}+{\rm{e}}^{-2 \beta E_{p}^{\pm}}, \\
U&=&G(\sigma^2+{\tilde{\Delta}}^2)+3G_8(\sigma^2+\tilde{\Delta}^2)^2\nonumber \\
&-&G_{\rm v}n_q^2,
\end{eqnarray}
where $N_{f}$ is the number of flavors, 
and $\tilde{\Delta} = |\langle q^TC i \gamma_5 \tau_2 t_2 q\rangle |$ is the diquark condensate. 
The $E_{p}^{\pm}$ are defined by 
\begin{eqnarray}
E_{p}^{\pm}={\rm sgn}(E_{p}\pm \tilde{{\mu}})\sqrt{(E_p \pm \tilde{{\mu}})^2+\Delta^2},  
\end{eqnarray}
where $E_p\equiv \sqrt{p^2+M^2}$ 
with the effective quark mass $M\equiv m-2G \sigma-4G_8 \sigma(\sigma^2+\tilde{\Delta}^2)$, 
$\tilde{{\mu}}=\mu-2G_{\rm v} n_q$, $\Delta =-2G\tilde{\Delta}$, and ${\rm sgn}(E_p \pm \tilde{{\mu}})$ is the sign function. 
When $m=\mu=0$, $\Omega$ becomes invariant under the rotation 
in $\sigma$--$\tilde{\Delta}$ plane as 
a consequence of Pauli-G{\rm{\"{u}}}rsey symmetry. 
Usually, $G$ and $G_8$ are assumed to be constant. 
However, they may depend on $\Phi$~\cite{Sakai_EPNJL}. 
Here we consider $\Phi$-dependent $G$ and $G_8$ defined by 
\bea 
G\equiv G_0(1-\alpha \Phi^2),~~~G_8\equiv G_{8,0}(1-\alpha \Phi^2), 
\label{G_G8_EPNJL}
\eea
where $G_0$, $G_{8,0}$ and $\alpha$ are constant parameters.  
In the Polyakov gauge, $\Phi$ is given by 
\begin{eqnarray}
\Phi=\frac{1}{2}({\rm{e}}^{i \phi}+{\rm{e}}^{-i \phi})=\cos(\phi),
\label{Phi}
\end{eqnarray}
for real number $\phi$. 
Following Ref.~\cite{Brauner}, we take the Polyakov-loop effective potential of the form 
\begin{eqnarray}
\frac{{\cal U}(\Phi)}{T}=-b\Big[24e^{-a/T}\Phi^2+\ln{(1-\Phi^2)}\Big].
\end{eqnarray}
As will be mentioned in the next section, we determine these parameters 
to reproduce LQC$_{2}$D data on $n_q$ at $\beta=0.75$ and 
the pseudo-critical temperature $T_{c0}=146$~MeV at $\mu =0$. 

Because this model is nonrenormalizable and the first two terms in $\Omega$ 
are divergent. 
We then regularize them by introducing a three-dimensional momentum cutoff as 
\begin{eqnarray}
\int \frac{d^3 p}{(2 \pi)^3} \rightarrow \frac{1}{2 \pi^2} \int_0^{\Lambda} dpp^2. 
\end{eqnarray}
The mean fields $X=\sigma, \Delta,n_q, \Phi$ are determined from the stationary conditions
\begin{eqnarray}
\frac{\partial \Omega}{\partial X}=0. 
\end{eqnarray}
We found that $\Delta$ is always zero at the temperature and the density we treat in this paper. 

There are six parameters in the NJL sector of the PNJL model. 
The values of $m,~G_{0},~G_{8,0}$ and $\Lambda$ are chosen 
to reproduce the vacuum property in LQC$_{2}$D simulations.    
The values of $G_{\rm v}$ and $\alpha$ are chosen to reproduce  
LQC$_{2}$D data on $n_q$; see Sec. \ref{ACP}. 
Table~\ref{para_PNJL_new} shows the parameter set used in this paper.

\begin{table}[htb]
\begin{center}
\begin{tabular}{c|c|c|c}
\hline\hline
$m_{\rm ps}[\rm{MeV}]$ & $M_{0}[\rm{MeV}]$ & $G_{0}[\rm{GeV}^{-2}]$ & $G_{8,0}[\rm{GeV}^{-8}]$  \\
\hline
616  & 354 & 4.6 & 60  \\
\hline \hline
 $G_{\rm v}/G_{0}$ & $\Lambda[{\rm{MeV}}]$ & $m[\rm{MeV}]$ & $\alpha$ \\
 \hline
 0.15 & 700& 110 & 0.2\\
 \hline \hline
\end{tabular}
\caption{Parameters of PNJL model. 
$M_{0}$ is the effective quark mass at vacuum.}
\label{para_PNJL_new}
\end{center}
\end{table} 

\section{Analytical continuation of physical quantities}
\label{ACP}

\subsection{\rm Analytical continuation}

In this section, we show the numerical results of LQC$_{2}$D simulations 
and perform the analytical continuation of physical quantities from the region 
at imaginary $\mu$ to the region at real $\mu$, and finally 
examine the validity of analytical continuation.

\begin{figure}[htb]
\begin{center}
\hspace{10pt}
\includegraphics[width=0.45\textwidth]{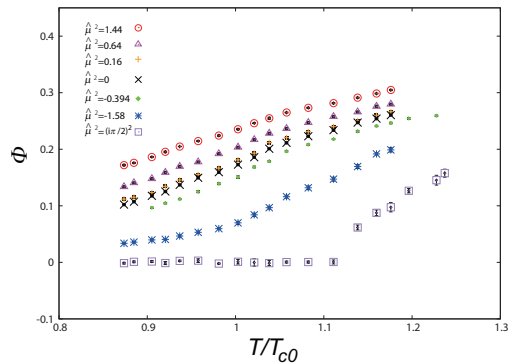}
\end{center}
\caption{LQC$_{2}$D results on $T$ dependence of $\Phi$ 
for several values of $\hat{\mu}^2$ 
}
\label{Pol_T}
\end{figure}

Figure~\ref{Pol_T} shows $T$ dependence of $\Phi$ 
for several values of $\hat{\mu} \equiv \mu/T$ from $i{\pi/{2}}$ to $1.2$. 
For all the cases except $\hat{\mu}^2=\left(i{\pi/{2}}\right)^2=-\left({\pi/{2}}\right)^2$, $\Phi$ increases smoothly as $T$ goes up. 
The deconfinement transition is thus crossover there. 
This property is the same as 
in QCD with three colors~\cite{YAoki_crossover}. 
For each $\hat{\mu}^2$, the pseudo-critical temperature $T_c(\hat{\mu}^2)$ 
(or $\beta_c(\hat{\mu}^2)$) is defined by the temperature 
where the susceptibility of $\Phi$ becomes maximum. 
It is found from LQC$_{2}$D simulations at $\hat{\mu } =0$ 
that $T_{c0} \equiv T_c(0) =146$~MeV. 
As for $\hat{\mu}^2=\left(i{\pi/{2}}\right)^2$, $\Phi$ is almost zero 
below $T=1.12T_{c0}$ and increases rapidly above $T=1.14T_{c0}$.  
This indicates that the transition is the second order. 
As mentioned in Sec.~\ref{sec:intro}, at $\hat{\mu} =i\theta =i{\pi/{2}}$ 
the system is symmetric under the ${\cal C}Z_2$ transformation, and 
$\Phi$ as an order parameter of spontaneous breaking of the symmetry is 
zero below $T_{\rm RW}$ and finite above $T_{\rm RW}$. Therefore, 
$T_{\rm RW}$ is located somewhere in a range of 
$T=1.12\sim 1.14T_{c0}$.  

\begin{figure}[htb]
\begin{center}
\hspace{10pt}
\includegraphics[width=0.445\textwidth,angle=0]{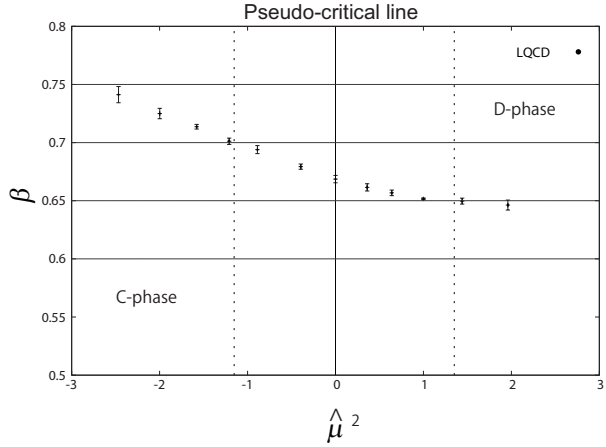}
\end{center}
\caption{Pseudo-critical line of deconfinemnet transition 
in $\hat{\mu}^2$--$\beta$ plane. 
In each of thin horizontal solid lines, $\beta$ is constant, 
and in each of thin vertical dotted lines $\hat{\mu}^2$ is constant . 
At $\beta =0.70$, the left hand side of the left thin vertical 
dotted line belongs to to the C-Phase, and the right side does to the D-phase. 
At $\beta =0.65$, the left hand side of the right thin vertical 
dotted line corresponds to the C-Phase, and the right side does to 
the D-phase. Note that $\beta_c(0)=0.67$.       
}
\label{phase_diagram_lqcd}
\end{figure}

Figure~\ref{phase_diagram_lqcd} shows 
the pseudo-critical line $\beta_c(\hat{\mu}^2)$ in $\hat{\mu}^2$--$\beta$ 
plane, and it is found that 
$\beta_c (\hat{\mu}^2)$ decreases as $\hat{\mu}^2$ increases. 
The value of $T_c(\hat{\mu}^2)$ at $\hat{\mu}^2=(i\pi/2)^2$ 
determined from $\beta_c((i\pi/2)^2)$ 
is $1.16T_{c0}$ and slightly larger than $T_{\rm RW}$ estimated above. 
This inconsistency is understood as follows. 
In our calculations, $\beta$ dependence of the Polyakov-loop susceptibility 
is calculated every 0.01 for each value of $\hat{\mu}^2$, 
and the data are fitted with a Gaussian function and the value of $\beta_c(\hat{\mu}^2)$ 
is evaluated as the point where the Gaussian function becomes maximum. 
The Gaussian fitting is, however, not valid in the vicinity of 
singular point (the endpoint of RW phase transition). 
We then adopt the value $1.12\sim 1.14T_{c0}$ as $T_{\rm RW}$.

Now we test the analytic continuation from imaginary $\mu$ to real $\mu$ for 
the cases of $\beta =0.75,0.70,0.65$ and $0.60$ that correspond to $T/T_{c0}=1.18, 1.06, 0.96, 0.88$, respectively. 
The system is in the deconfinement (D) phase at $\beta =0.75$, 
while it is in the confinement (C) phase at $\beta =0.60$. 
At $\beta =0.70~(0.65)$, the system is in the C-phase when $\hat{\mu}^2<-1.15~(1.35)$ and 
in the D-phase when $\hat{\mu}^2>-1.15~(1.35)$. 
For each temperature, we then use different analytic functions 
as explained below~\cite{Cea_1,Cea_2,Lombardo_2005}. 

\subsubsection{$T_{\rm RW}<T(\beta=0.75)$}

At this temperature, due to the existence of RW transition, 
physical quantities cannot be described by any continuous periodic function. 
Hence we use a polynomial series of the form 
\bea
A+B\hat{\mu}^2,
\label{function_pol_chiral_high} 
\eea
or
\begin{eqnarray}
A+B\hat{\mu}^2+C\hat{\mu}^4,
\label{function_pol_chiral_high_4}
\end{eqnarray}
for $\hat{\mu}$-even quantities $\Phi$ and $\sigma$, 
where $A$, $B$, $C$ are expansion coefficient.     
For a $\hat{\mu}$-odd quantity $n_q$, we use 
\begin{eqnarray}
A\hat{\mu}+B\hat{\mu}^3,
\label{function_density_high}
\end{eqnarray}
or 
\begin{eqnarray}
A\hat{\mu}+B\hat{\mu}^3+C\hat{\mu}^5. 
\label{function_density_high_5}
\end{eqnarray}

\subsubsection{$T_{c0}<T(\beta=0.70)<T_{\rm RW}$}

At this temperature, the system is in the D-phase when $\hat{\mu}^2>-1.15$. 
We then use the same polynomial series as in the case of $\beta =0.75$, 
but consider only the region $-1.15<\hat{\mu}^2\le 0$ as a fitting range.
For $\Phi$ and $\sigma$, 
we use only the quadratic function (\ref{function_pol_chiral_high}), 
since the number of data we can use is small.

\subsubsection{$T(\beta=0.60,~0.65)<T_{c0}$}

At this temperature, the system is in the C-phase at imaginary and zero 
$\hat{\mu}$. Hence, it is expected that physical quantities can be well 
described by  continuous periodic functions. 
Since $\Phi (\theta )$ is $\theta$-even and has a periodicity of $2\pi$ 
in $\theta ={\rm Im}(\hat{\mu})$, we use the following Fourier series 
\begin{eqnarray}
A\cos(\theta ), 
\label{function_pol_low}
\end{eqnarray}
or 
\begin{eqnarray}
A\cos(\theta ) +B \cos(3\theta ),  
\label{function_pol_low_3}
\end{eqnarray}
for $\Phi$. 
Note that the terms of $\cos(2\theta )$ and $\cos(4\theta )$ 
as well as the constant term are excluded, since 
$\Phi ({\pi/{2}}+\theta^\prime )=-\Phi ({\pi/{2}}-\theta^\prime )$ 
for any $\theta^\prime$.  
The chiral condensate $\sigma (\theta )$ is a $\theta$-even and periodic 
function with a period $\pi$. 
We then use the following Fourier series 
\begin{eqnarray}
A+B \cos(2\theta ), 
\label{function_chiral_low}
\end{eqnarray}
or 
\begin{eqnarray}
A+B \cos(2\theta )+ C\cos(4\theta ), 
\label{function_chiral_low_4}
\end{eqnarray}
for $\sigma$. 
The quark number density $n_q(\theta )$ is a $\theta$-odd and periodic 
function with a period $\pi$. We therefore use the following Fourier series
\begin{eqnarray}
A\sin(2\theta ), 
\label{function_density_low}
\end{eqnarray}
or 
\begin{eqnarray}
A\sin(2\theta ) +B \sin(4\theta ), 
\label{function_density_low_4}
\end{eqnarray}
for $n_q$. 
Note that, in the case of $\beta =0.65$, the system is in the D-phase when $\hat{\mu}^2>1.35$. 
Hence, the Fourier series in which the coefficients are determined from 
LQCD  data at imaginary $\hat{\mu}$ and zero $\hat{\mu}=0$ may not work there.

\subsubsection{Pseudo-critical line}

The pseudo-critical line $\beta_{c}(\hat{\mu}^2)$ is $\hat{\mu}$-even. 
We then use the polynomial series (\ref{function_pol_chiral_high}) and (\ref{function_pol_chiral_high_4}).

\subsection{Quark number density}
\label{Sec:quark number density}

\begin{figure}[htb]
\begin{center}
\hspace{10pt}
\includegraphics[angle=0,width=0.4\textwidth]{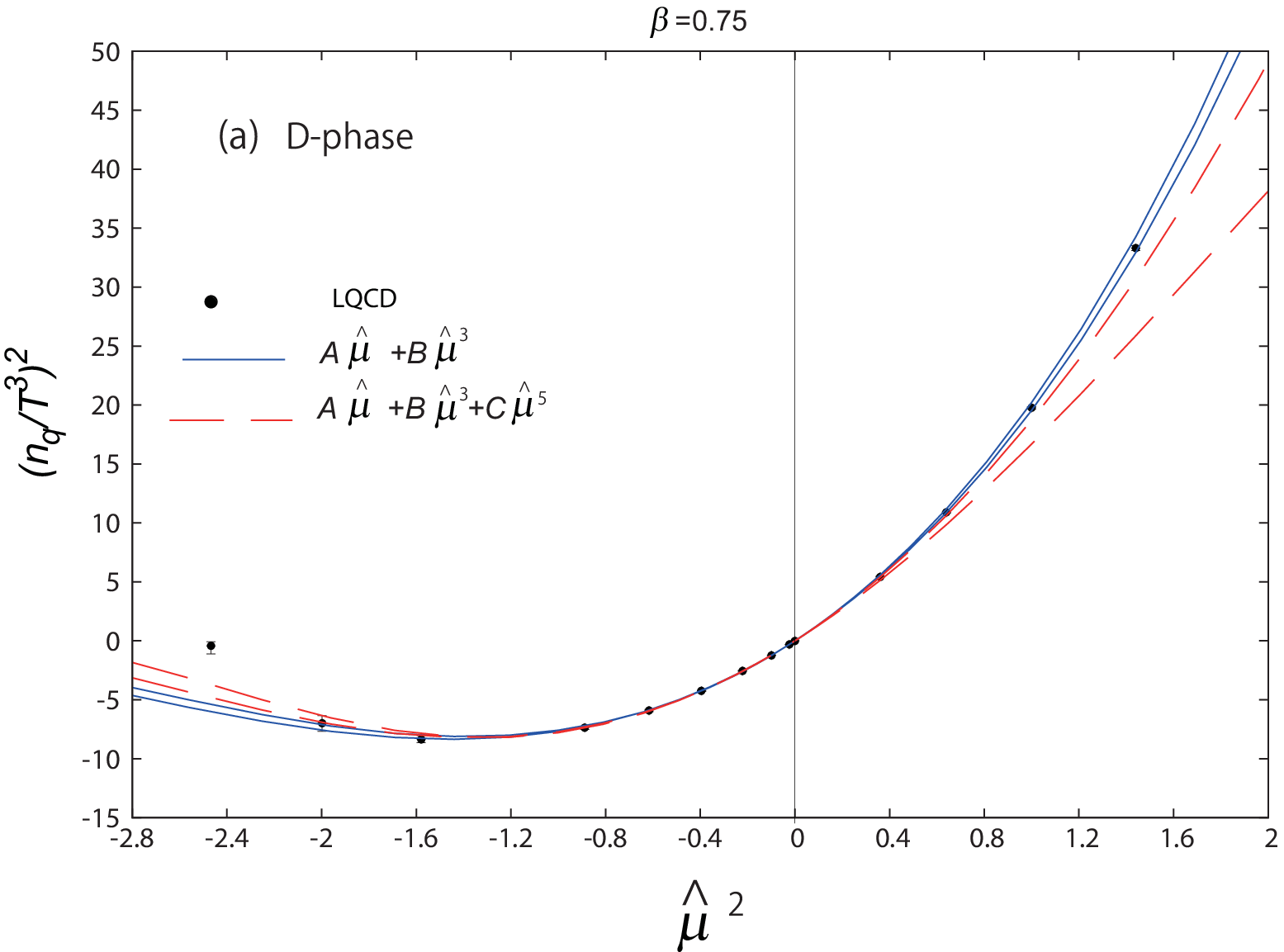}
\includegraphics[angle=0,width=0.4\textwidth]{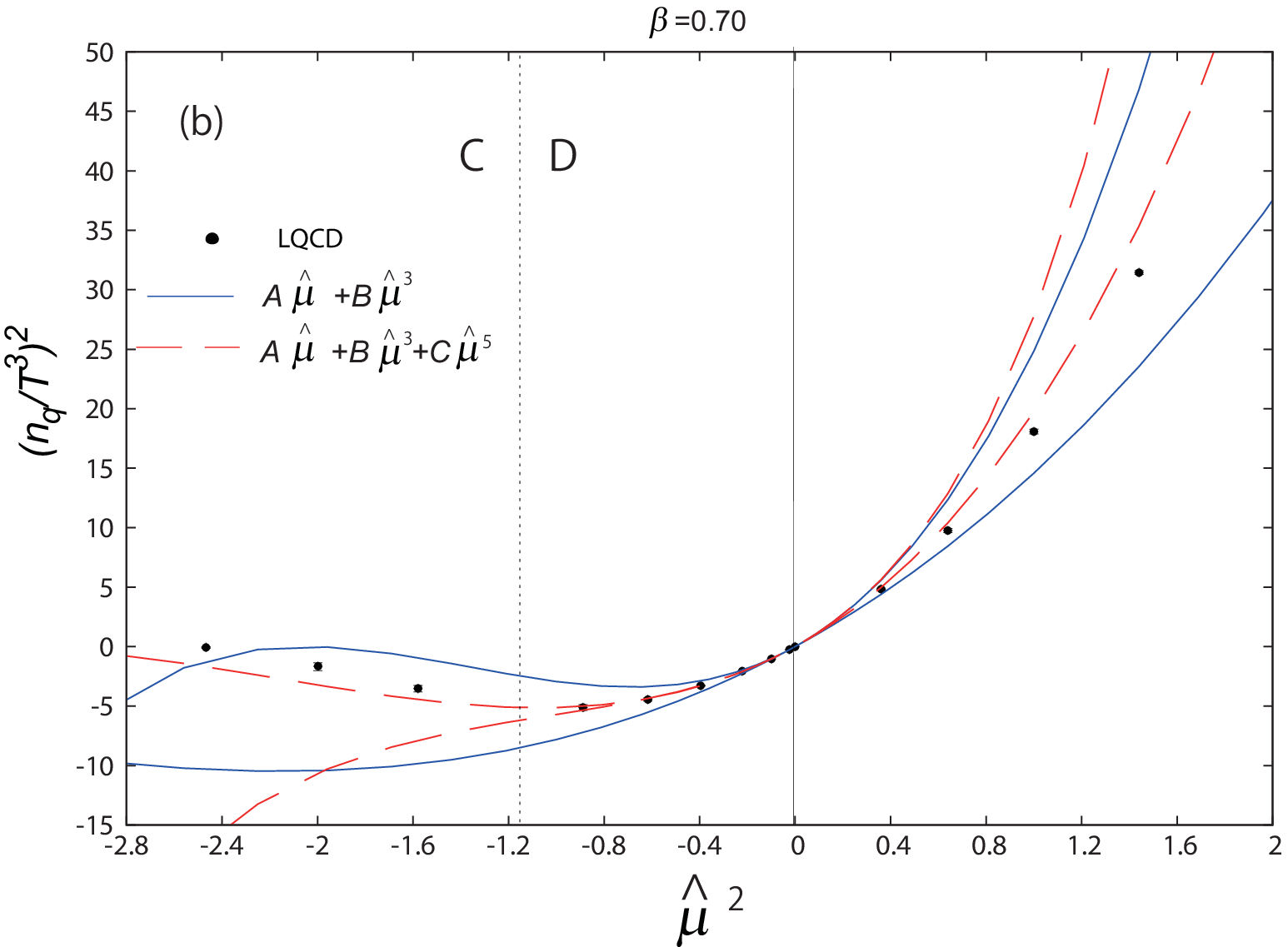}
\includegraphics[angle=0,width=0.4\textwidth]{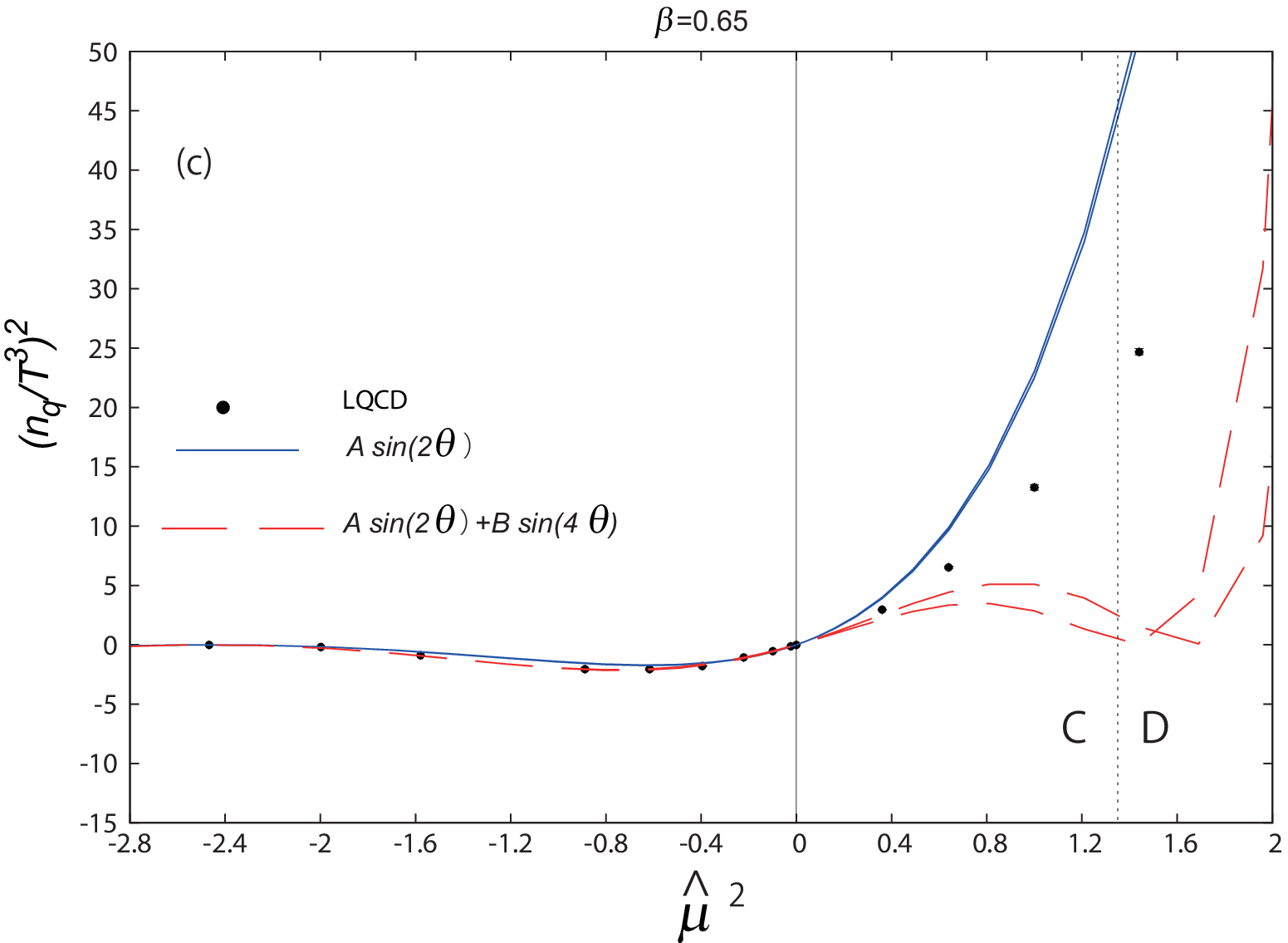}
\includegraphics[angle=0,width=0.4\textwidth]{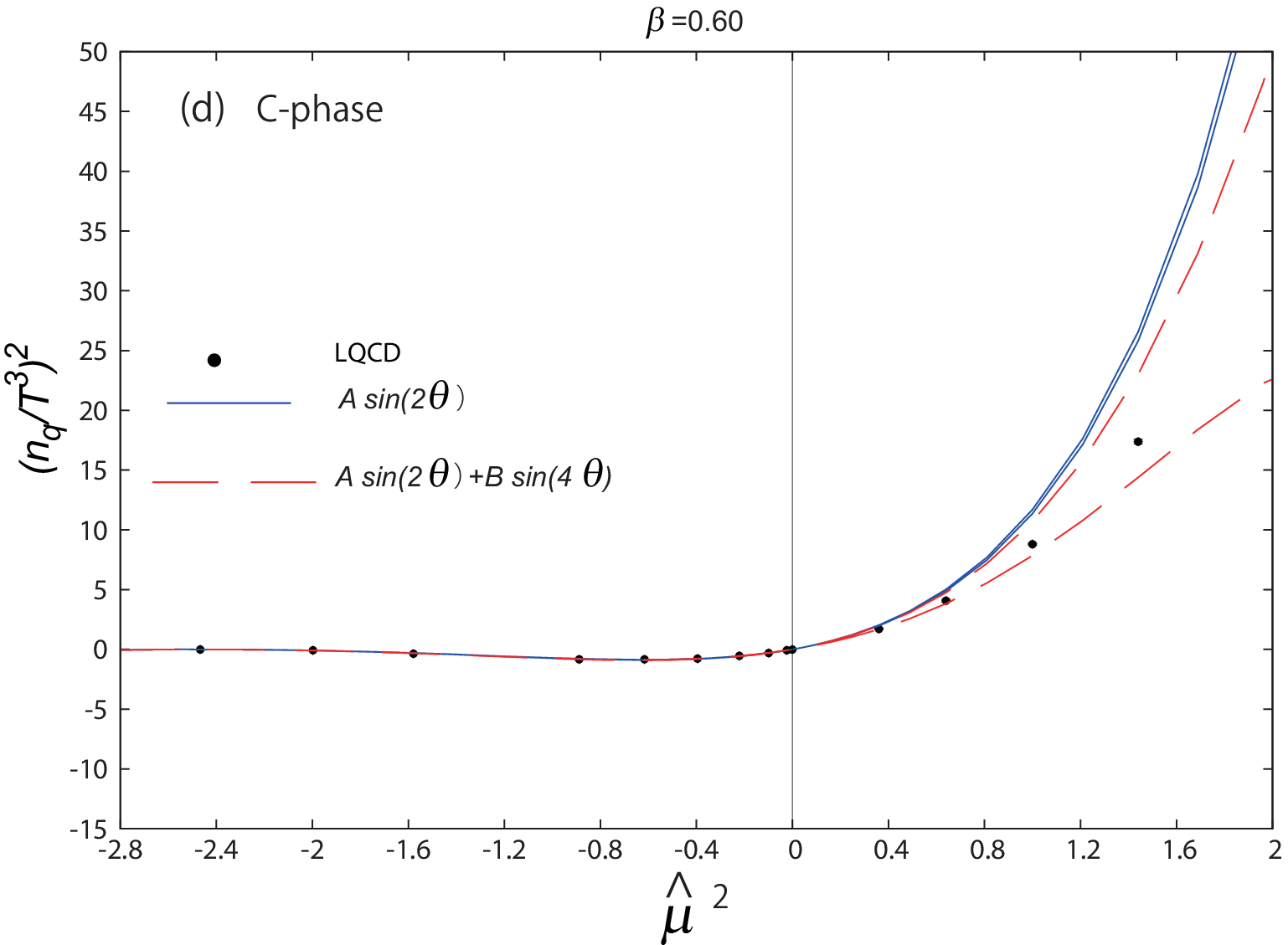}
\end{center}
\caption{
$\hat{\mu}^2$ dependence of $n_q$ at (a) $\beta=0.75~(T/T_{c0}=1.18)$, 
(b) $\beta = 0.70~(T/T_{c0}=1.06)$, (c)  $\beta =0.65~(T/T_{c0}=0.96)$ and 
(d) $\beta =0.60~(T/T_{c0}=0.88)$.  
The dots with error bars are the results of LQC$_2$D data. 
The solid and dashed lines represent the results of analytic continuation 
in which two types of analytic functions are taken as shown by 
legends. 
The upper and lower bounds of analytic continuation is shown by 
a pair of same lines. 
Characters C and D denote confinement and deconfinement phases, 
respectively. 
}
\label{nq_muT_1}
\end{figure}

First we consider the analytic continuation of $n_q$.  
Figure~\ref{nq_muT_1} shows $\hat{\mu}^2$ dependence of $(n_q/T)^2$ 
for several 
values of $T$. The analytic continuation has errors coming from LQCD data at 
zero and imaginary $\hat{\mu}$. We then plot the upper and lower bounds of 
analytic continuation by a pair of same lines. 
The $(n_q/T)^2$ are smooth at $\hat{\mu}=0$, as expected. 
This is true for $\delta \sigma$ and $\Phi$, as shown later. 
This guarantees that the analytic continuation 
from imaginary $\hat{\mu}$ to real $\hat{\mu}$ is possible.

At $\beta =0.75~(T/T_{c0}=1.18)$, the system is in the D-phase 
and hence the polynomial series is used. 
The coefficients determined from LQCD data at imaginary $\mu$ are tabulated 
in Table~\ref{para_quantity}~(a) of Appendix 
\ref{Parameters of fitting functions}. 
The polynomial series up to $\hat{\mu}^3$ well reproduces LQC$_2$D data 
in a wide range of $0\leq \hat{\mu}^2 \leq (1.2)^2$. 
Note that the analytic function deviates from LQC$_2$D data 
near the first-order RW phase transition present at 
$T > T_{\rm RW}$ and $\hat{\mu}^2=-(\pi/2)^2$.

At $\beta =0.70~(T/T_{c0}=1.06)$, the system is in the C-phase when $\hat{\mu}^2<-1.15$, while it is in the D-phase otherwise. 
Hence, we use only seven data in a range of $\hat{\mu}^2=-1.15\sim 0$ 
to determine the coefficients of polynomial series.  
The coefficients of the function are tabulated 
in Table~\ref{para_quantity}~(b) of Appendix 
\ref{Parameters of fitting functions}. 
LQC$_2$D data calculated at real $\hat{\mu}$ 
lie between the upper and the lower bounds of the polynomial series up to $\hat{\mu}^3$ in a wide range of $0\leq \hat{\mu}^2 \leq (1.2)^2$.

At $\beta =0.65~(T/T_{c0}=0.96)$, the system is in the C-phase when $\hat{\mu}^2<1.35$, while it is in the D-phase otherwise.  
Hence, we use the Fourier series.  
The coefficients of the function are tabulated 
in Table~\ref{para_quantity}~(c) of Appendix 
\ref{Parameters of fitting functions}.  
The analytic functions fail to reproduce LQC$_2$D data calculated at 
real $\hat{\mu}$ when $\hat{\mu}^2\ge 0.4$.   
The large deviation at large $\hat{\mu}^2$ may be originated in the fact that 
the system is in the D-phase there and the Fourier series may not be valid 
anymore.    

At $\beta =0.60~(T/T_{c0}=0.88)$, the system is in the C-phase.   
Therefore, we use the Fourier series.  
The coefficients of the function are tabulated in 
Table~\ref{para_quantity}~(d) of Appendix 
\ref{Parameters of fitting functions}.   
The Fourier series up to the term $\sin{(2\theta )}$ ($\sin{(4\theta )}$) are 
consistent with LQC$_2$D data calculated at real $\hat{\mu}$ 
in a wide range $0\le \hat{\mu}^2< 0.8$ ($0\le \hat{\mu}^2 \leq (1.2)^2$).      

Comparing four cases of $T/T_{c0}$ with each other, 
one can see that the analytic continuation is reasonable 
at higher $T/T_{c0}$ where the system is always in the D-phase 
when $\hat{\mu}^2$ varies from $-(\pi/2)^2$ to $(1.2)^2$
and at lower $T/T_{c0}$ where the system is always in the C-phase 
when $\hat{\mu}^2$ varies from $-(\pi/2)^2$ to $(1.2)^2$. 
Near $T/T_{c0}=1$, however, the system changes from the C-phase to 
the D-phase as $\hat{\mu}^2$ varies from $-(\pi/2)^2$ to a positive value. 
A simple analytic function cannot follow the complicated change properly. 
Therefore, the analytic continuation is reasonable 
except for the vicinity of deconfinement crossover.

\subsection{Chiral condensate}
\label{Sec:chiralcondensate}

\begin{figure}[htb]
\begin{center}
\hspace{10pt}
\includegraphics[angle=0,width=0.4\textwidth]{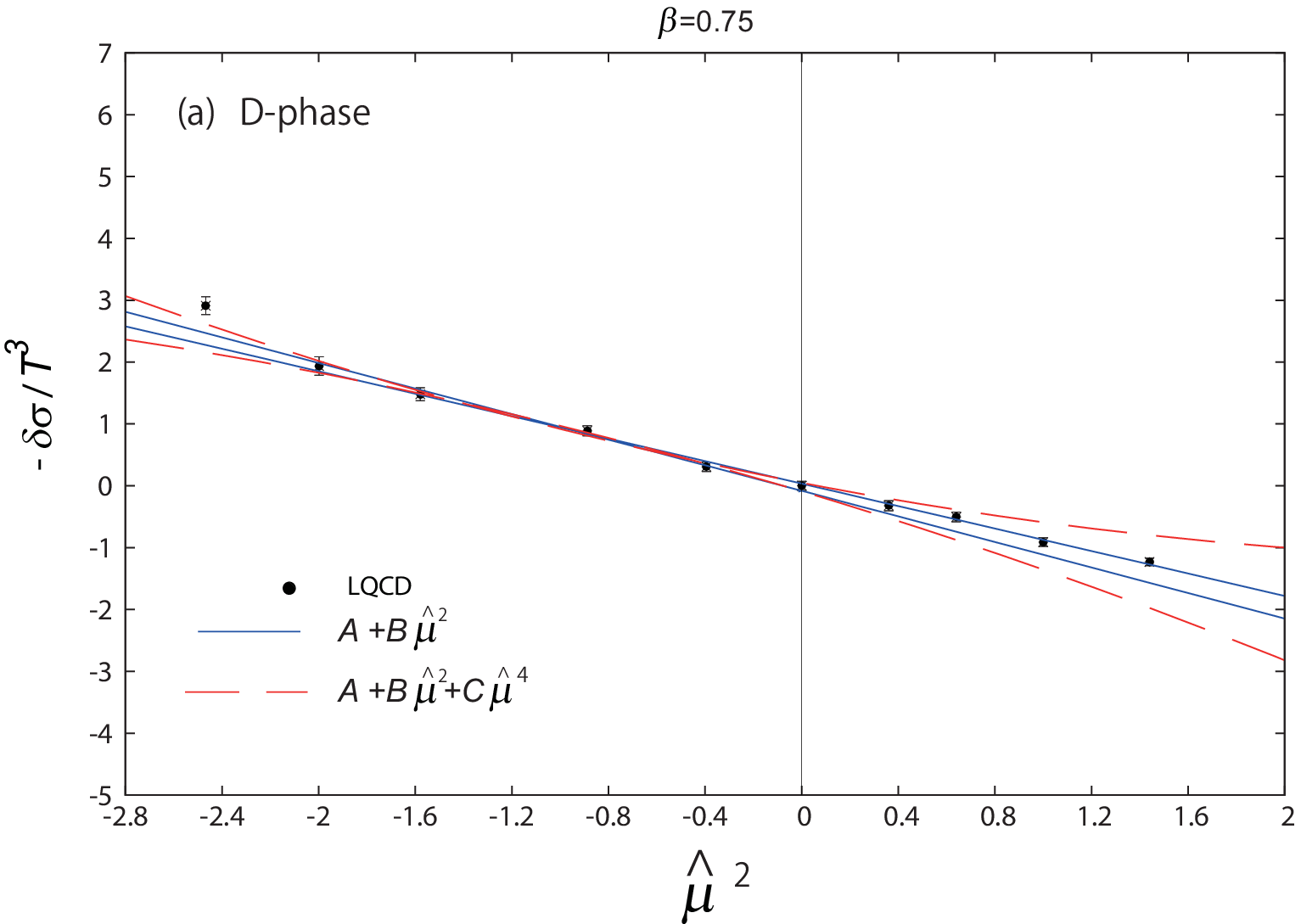}
\includegraphics[angle=0,width=0.4\textwidth]{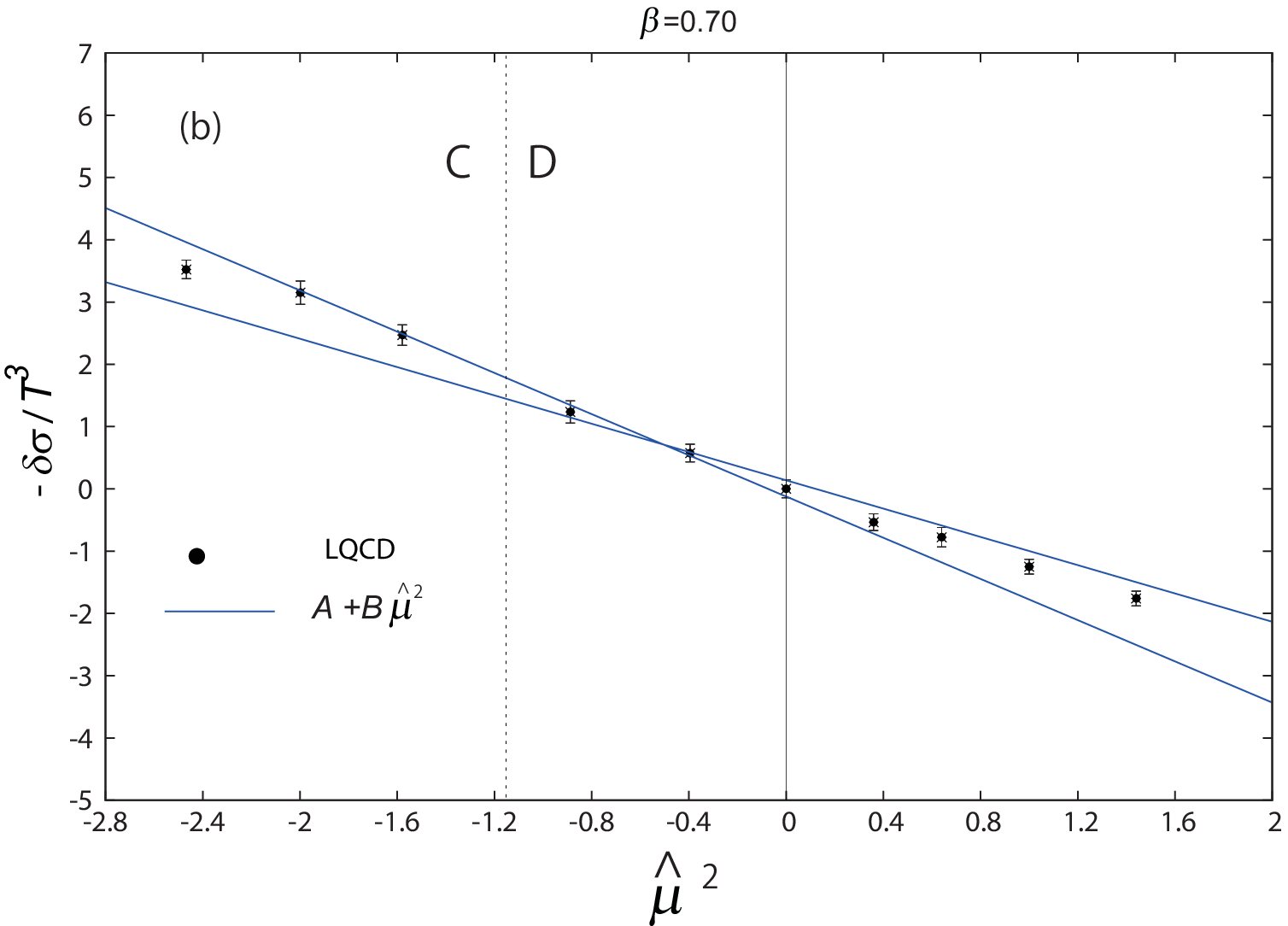}
\includegraphics[angle=0,width=0.4\textwidth]{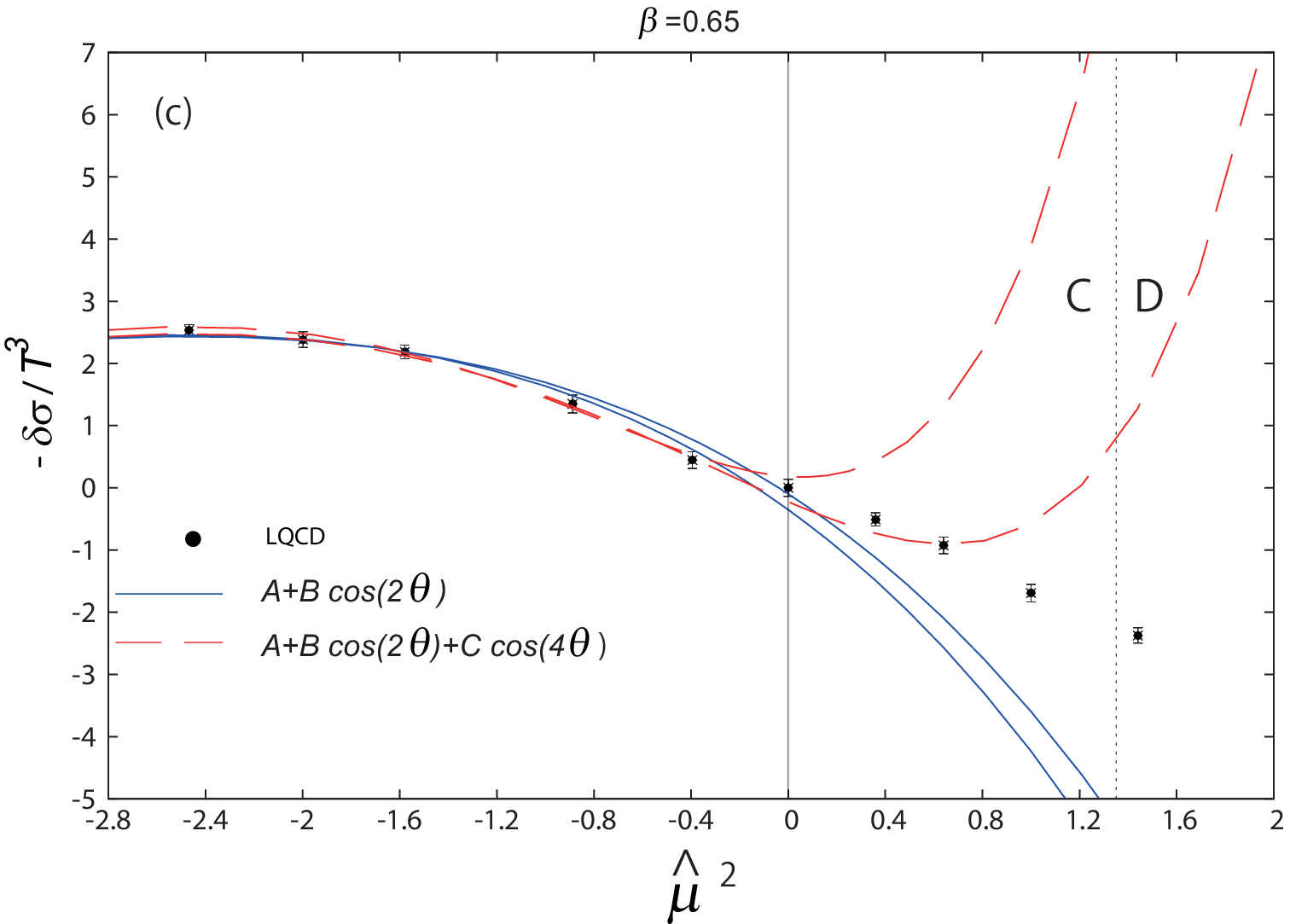}
\includegraphics[angle=0,width=0.4\textwidth]{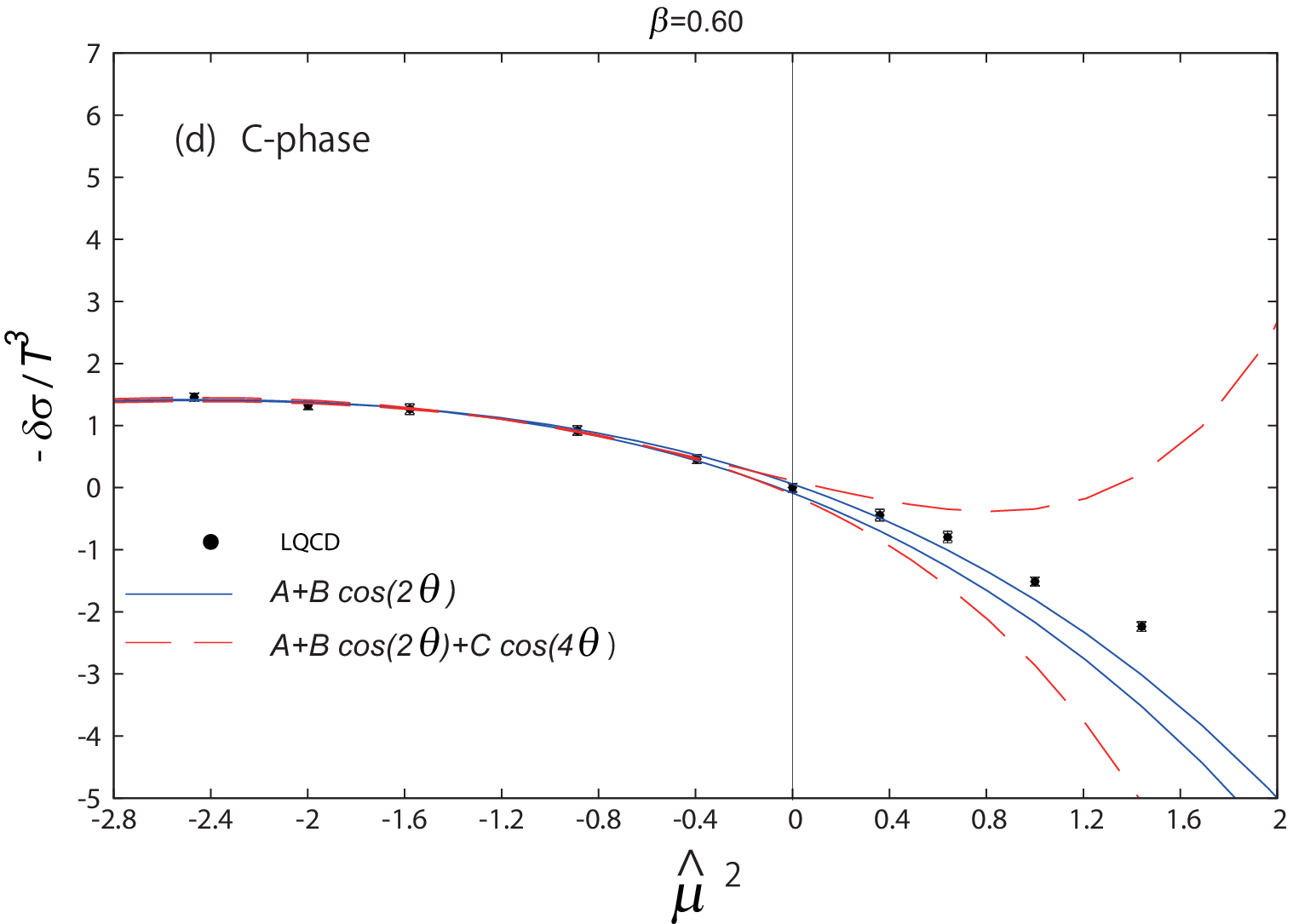}
\end{center}
\caption{
$\hat{\mu}^2$-dependence of $\sigma$ at (a) $\beta=0.75~(T/T_{c0}=1.18)$ (b) $\beta = 0.70~(T/T_{c0}=1.06)$ (c)  $\beta =0.65~(T/T_{c0}=0.96)$
(d) $\beta =0.60~(T/T_{c0}=0.88)$. 
For the definition of dots, lines and characters, 
see the caption of Fig.~\ref{nq_muT_1}.
}
\label{chiral_muT_1}
\end{figure}

Figure~\ref{chiral_muT_1} shows $\hat{\mu}^2$ dependence of $\delta{\sigma}$ 
for several values of $T$. Again, the upper and lower bounds 
of analytic continuation are shown by a pair of same lines; 
see Table~\ref{para_quantity} for the coefficients of analytic function 
determined from LQC$_2$D data at imaginary $\mu$. 
As for $\delta{\sigma}$, one can made the same discussion as 
in Sec. \ref{Sec:quark number density} for $n_q$, 
as shown 
below, although the analytic function taken in the C-phase is 
a cosine function.

At $\beta =0.75~(T/T_{c0}=1.18)$, the polynomial series 
up to $\hat{\mu}^2$ well reproduces LQC$_2$D data 
calculated at real $\hat{\mu}$ in a wide range of 
$0\leq \hat{\mu}^2 \leq (1.2)^2$. 
At $\beta =0.70~(T/T_{c0}=1.06)$, the system is in the C-phase when $\hat{\mu}^2<-1.15$, while it is in the D-phase otherwise.  
Hence, we can use only three data in a range of $\hat{\mu}^2=-1.15\sim 0$ 
to determine the coefficients of analytic function and then 
use the quadratic function only.  The function is 
consistent with LQC$_2$D calculated at real $\hat{\mu}$
in a wide range of $0\leq \hat{\mu}^2 \leq (1.2)^2$. 
  
At $\beta =0.65~(T/T_{c0}=0.96)$, the system is in the C-phase when $\hat{\mu}^2<1.35$, while it is in the D-phase otherwise.  
Hence, we use the Fourier series.  
The analytic functions is not consistent with LQC$_2$D data 
calculated at real $\hat{\mu}$ when $\hat{\mu}^2>0.4$. 
As mentioned in the case of $n_q$, this failure at large $\hat{\mu}$ may 
show that the system is in the D-phase there and 
the Fourier series becomes less reliable. 
At $\beta =0.60~(T/T_{c0}=0.88)$, the system is in the C-phase.   
Hence, we use the Fourier series.  
The analytic functions are consistent with LQC$_2$D data calculated 
at real $\hat{\mu}$ in a range of $0\leq \hat{\mu}^2 <0.8$

\subsection{\rm{Polyakov loop}}

\begin{figure}[htb]
\begin{center}
\hspace{10pt}
\includegraphics[angle=0,width=0.4\textwidth]{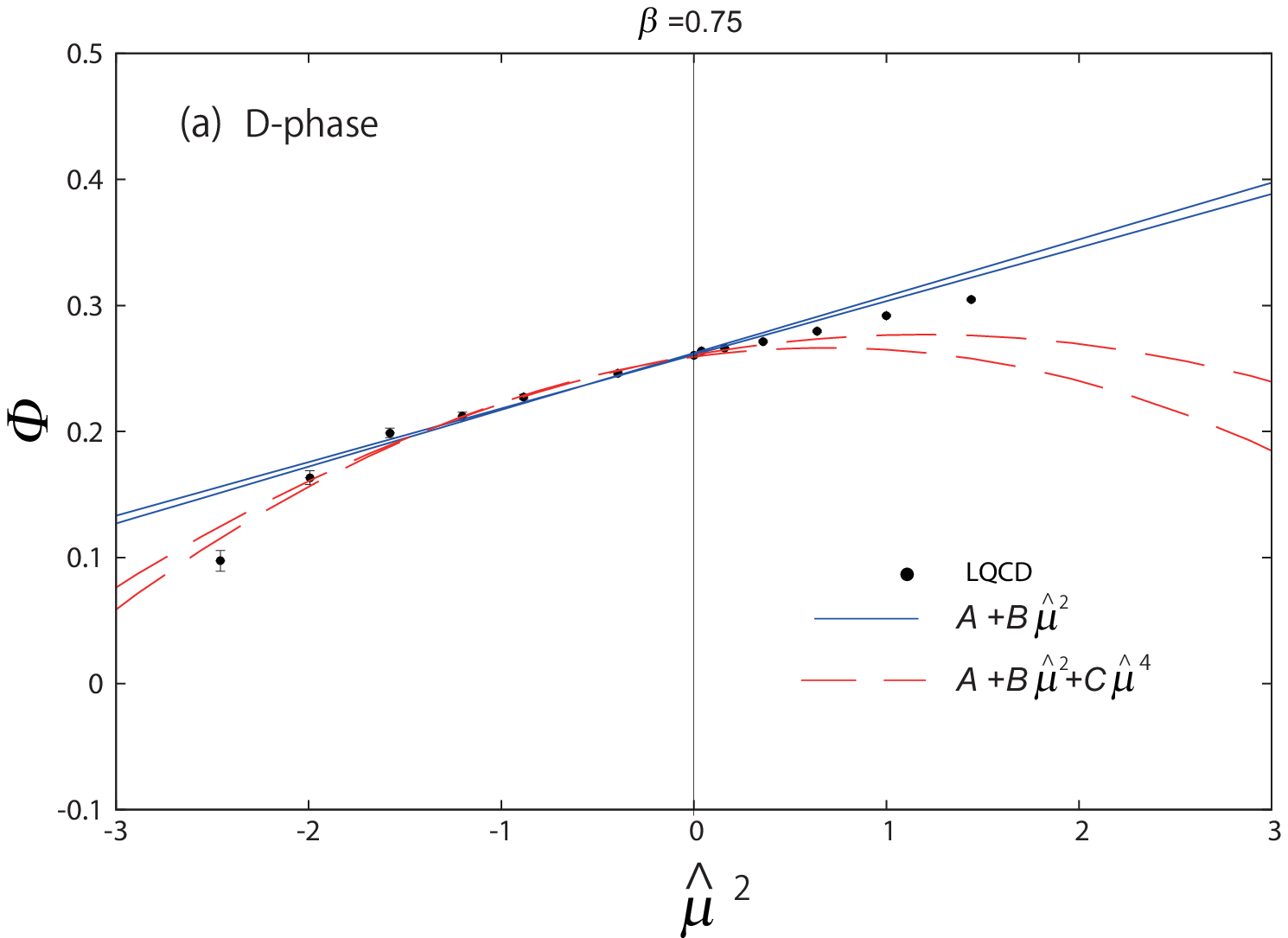}
\includegraphics[angle=0,width=0.4\textwidth]{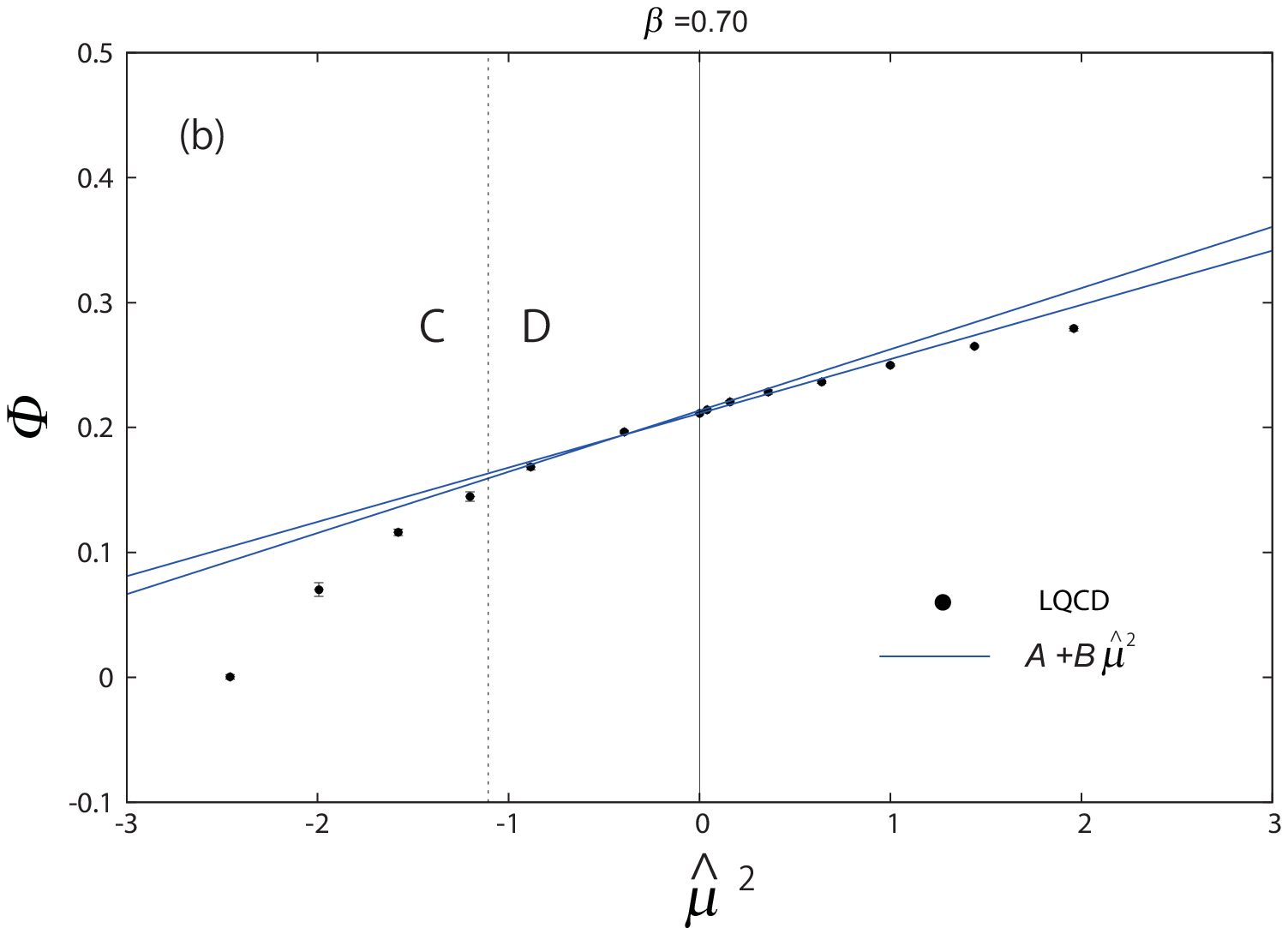}
\includegraphics[angle=0,width=0.4\textwidth]{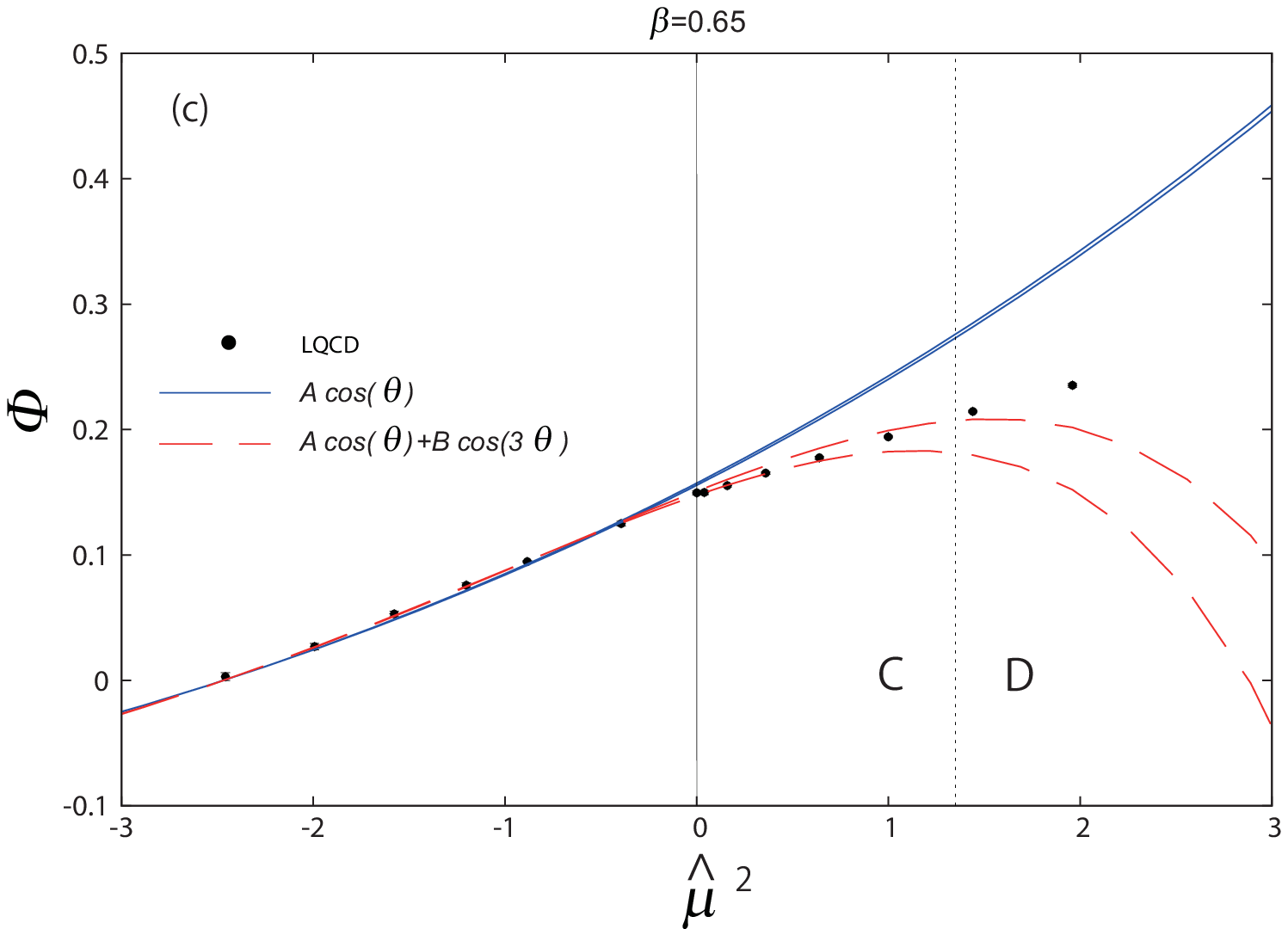}
\includegraphics[angle=0,width=0.4\textwidth]{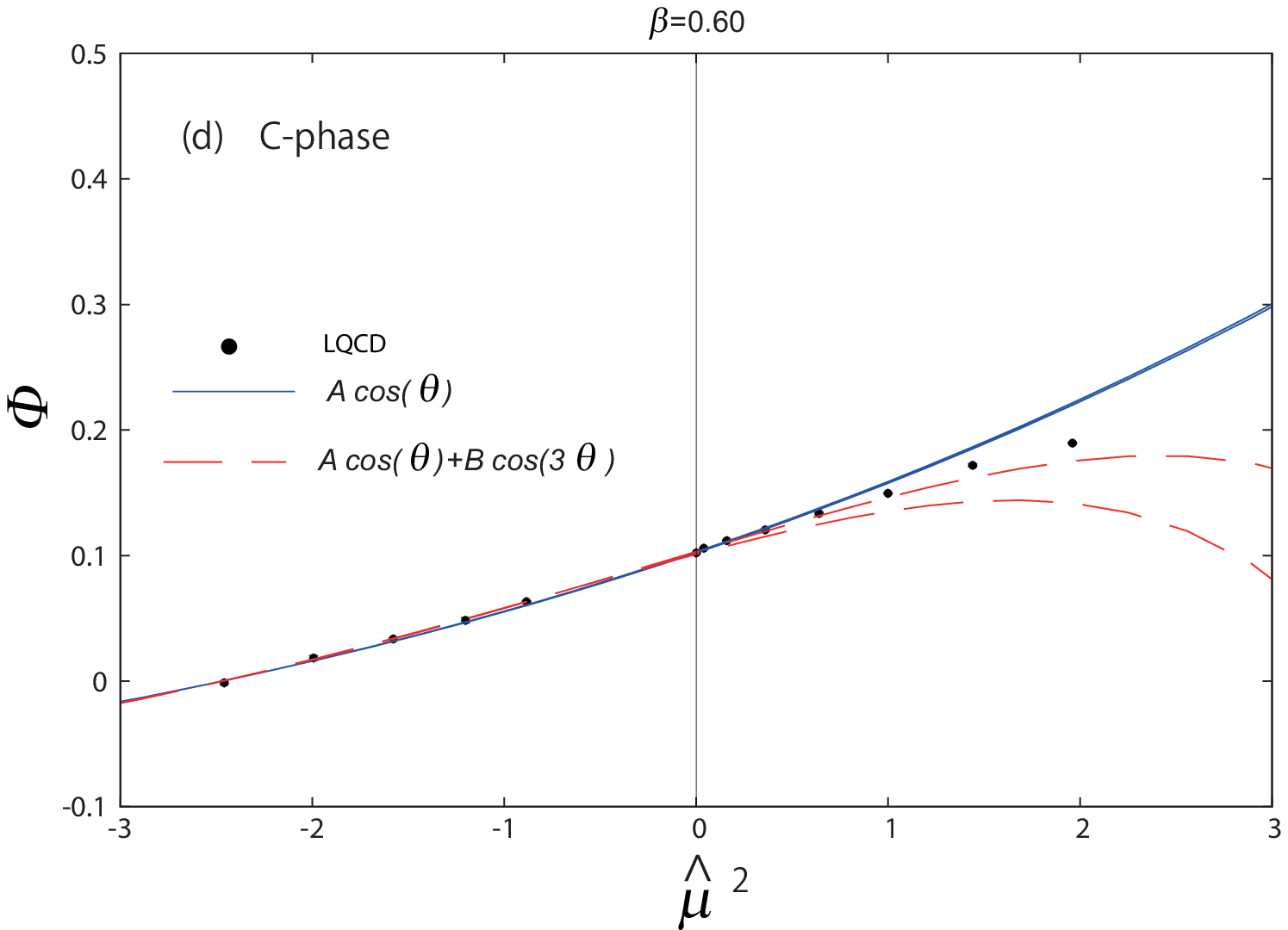}
\end{center}
\caption{
$\hat{\mu}^2$-dependence of $\Phi$ at (a) $\beta=0.75~(T/T_{c0}=1.18)$, 
(b) $\beta = 0.70~(T/T_{c0}=1.06)$, 
(c)  $\beta =0.65~(T/T_{c0}=0.96)$ and 
(d) $\beta =0.60~(T/T_{c0}=0.88)$. 
For the definition of dots, lines and characters, see the caption 
of Fig.~\ref{nq_muT_1}.
}  
\label{Pol_muT2_1}
\end{figure}

Figure~\ref{Pol_muT2_1} shows $\hat{\mu}^2$ dependence of $\Phi$ 
at several values of $T$. Again, the upper and the lower bounds of 
analytic continuation are shown by a pair of same lines; 
see Table~\ref{para_quantity} for the coefficients of analytic function. 
As for $\Phi$, one can make the same discussion qualitatively 
as in Sec. \ref{Sec:chiralcondensate} for $\delta{\sigma}$.

\subsection{Pseudo-critical line}
\label{Sec:Phasedigram}

\begin{figure}[htb]
\begin{center}
\hspace{10pt}
\includegraphics[angle=0,width=0.445\textwidth]{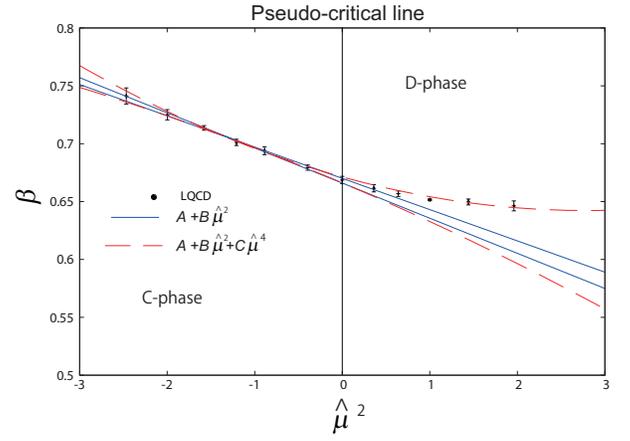}
\end{center}
\caption{
Pseudo-critical line of deconfinement crossover in $\hat{\mu}^2$--$\beta$ plane. 
Note that $\beta_c(0)=0.67$. 
For the definition of dots, lines and characters, 
see the caption of Fig.~\ref{nq_muT_1}.  
}
\label{phase_diagram}
\end{figure}

Figure~\ref{phase_diagram} shows the transition line of deconfinement 
crossover in $\hat{\mu}^2$--$\beta$ plane. 
The pseudo-critical $\beta_c(\hat{\mu}^2)$ at $\hat{\mu}^2=0$ is about 0.67.   
We consider the polynomial series the coefficients of which  
are obtained from LQC$_2$D data at imaginary $\mu$ and tabulated 
in Table~\ref{para_phase} of Appendix \ref{Parameters of fitting functions}. 
The polynomial series up to $\hat{\mu}^2$ well reproduces LQC$_2$D data 
at $0\leq  \hat{\mu}^2 <0.8$, but deviates at $\hat{\mu}^2 >0.8$. 
The polynomial series up to $\hat{\mu}^4$  is consistent with LQC$_2$D data 
even at $\hat{\mu}^2 >0.8$, but the difference between 
the upper and lower bounds of analytic continuation is large.  
Therefore, we should consider that the analytic continuation of the pseudo-critical line is reasonable at $0\leq  \hat{\mu}^2 <0.8$.

\clearpage

\section{Comparison of PNJL model with LQC$_2$D data}
\label{ComLP}
\subsection{\rm Parameter setting}
\label{Sec:comparison}

In this section, we compare results of the PNJL model with LQC$_2$D data  
to test the validity of the model. 
For this purpose, we first fix the parameters of the model. 
For the NJL sector, the parameters have already been determined  
in Sec. \ref{PNJLM}. We then fix the remaining parameters 
$a$, $b$, $\alpha$ and $G_{\rm v}$ here.

\begin{figure}[htb]
\begin{center}
\hspace{10pt}
\includegraphics[width=0.445\textwidth,angle=0]{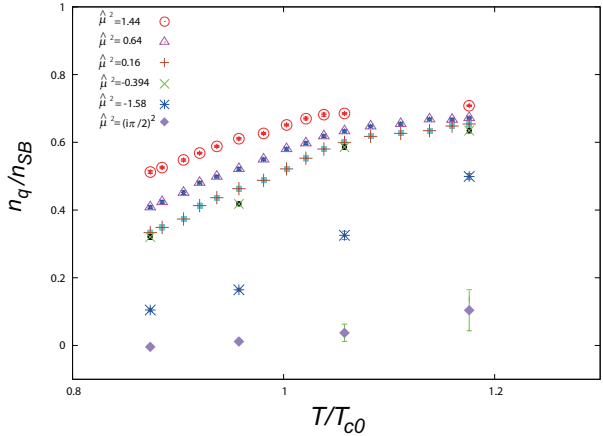}
\end{center}

\caption{$T$ dependence of $n_q/n_{\rm SB}$ for several values of 
$\hat{\mu}^2$. Six cases of $\hat{\mu}^2 =\left(i{\pi\over{2}}\right)^2, -1.58, -0.394, 0.16, 0.64, 1.44$ are shown from the bottom. 
}
\label{Density_T}
\end{figure}

Figure~\ref{Density_T} shows $T$ dependence of $n_q$ divided by its Stephan-Boltzmann (SB) limit $n_{\rm SB}$ for several values of $\hat{\mu}^2$ from 
$-(\pi/2)^2$ to $(1.2)^2$. LQC$_2$D results include a lattice artifact 
due to finite volume and spacing. 
The artifact is expected to be reduced in $n_q/n_{\rm SB}$. 
For all the values of $\hat{\mu}^2$, 
the ratio $n_q/n_{\rm SB}$ increases as $T$ increases.

Now we determine the parameters $a$, $b$, $\alpha$ and $G_{\rm v}$ 
from $n_q$ at the highest $T$ 
in the present analyses, i.e., at $\beta=0.75$ ($T/T_{c0}=1.18$). 
One reason is that 
the PNJL model is essentially a model for quark dynamics and 
it may work better at higher $T$ than at lower $T$. 
Another reason is that $n_q$ does not need the renormalization and is 
sensitive to the value of $G_{\rm v}$.

\begin{figure}[htb]
\begin{center}
\hspace{10pt}
\includegraphics[angle=0,width=0.445\textwidth]{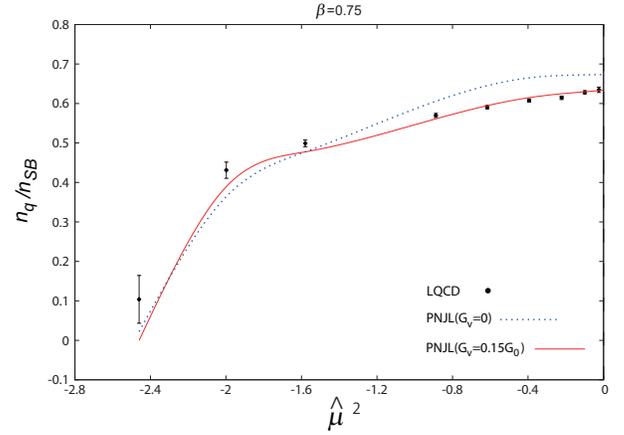}
\end{center}
\caption{$\hat{\mu}^2$ dependence of $n_q/n_{\rm SB}$ 
at $\beta=0.75$ at imaginary $\mu$.  
The dots with error bars are the results of LQC$_2$D data. 
The solid (dotted) line is the result of PNJL model 
with $G_{\rm v}/G_0=0.15~(0)$. 
}
\label{compare_n_beta_0.75_im}
\end{figure}

Figure~\ref{compare_n_beta_0.75_im} shows $\hat{\mu}^2$ dependence of 
$n_q/n_{\rm SB}$ at $\beta =0.75$ at imaginary $\mu$. 
As $\hat{\mu}^2$ decreases, the chiral symmetry breaking becomes stronger and 
the effective quark mass $M$ becomes larger~\cite{Sakai_IM}, so that 
$n_q/n_{\rm SB}$ decreases. 
We searched the parameters $a$, $b$, $\alpha$ and $G_{\rm v}$ so as to 
reproduce both the result of Fig.~\ref{compare_n_beta_0.75_im} and 
$T_{c0}=146$~MeV. 
The parameters thus obtained are shown 
in Table~\ref{para_PNJL_new} of Sec.~\ref{PNJLM}. 
It is interesting that the vector interaction is needed to reproduce 
LQC$_2$D results  data at imaginary $\mu$. 
In fact, the model with $G_{\rm v}=0.15G_{0}$ (solid line) yields 
better agreement with 
LQC$_2$D data than the model with $G_{\rm v}=0$ (dotted line). 
This method may work as a way of 
determining the vector coupling also in realistic QCD 
with three colors~\cite{Sakai_vector,Sakai_para,Kashiwa_nonlocal}.  

\begin{figure}[htb]
\begin{center}
\hspace{10pt}
\includegraphics[angle=0,width=0.445\textwidth]{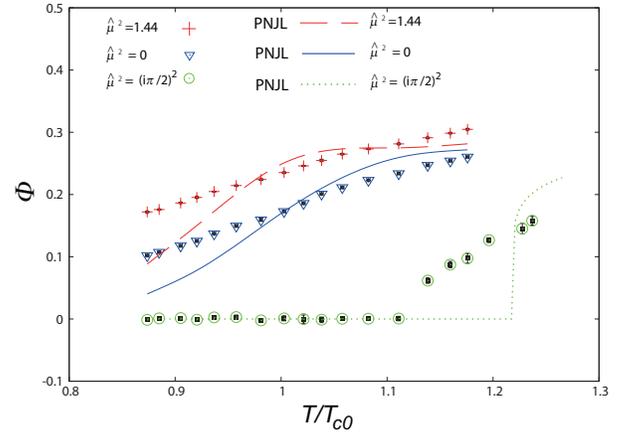}
\end{center}
\caption{
$T$ dependence of $\Phi$ for three values of $\hat{\mu}^2$. 
The dots represent LQC$_2$D data. 
The solid (dotted, dashed) line represents the PNJL results at  
$\hat{\mu}^2 =0~(-\left({\pi/{2}}\right)^2), 1.44)$. 
The PNJL results are multiplied by the normalization factor 0.304.}  
\label{Pol_TTc_1}
\end{figure}

$T$ dependence of $\Phi$ is shown in Fig. ~\ref{Pol_TTc_1} for 
three cases of $\hat{\mu}^2=-(\pi/2)^2, 0, 1.44$. 
Since the renormalization is needed for $\Phi$, 
we multiply the PNJL results by a factor 0.304 to reproduce 
LQC$_2$D results at $\hat{\mu}=0$ and $T=T_{c0}$. 
The renormalized PNJL results (solid lines) reproduce LQC$_2$D data 
qualitatively, except for the vicinity of the first-order RW phase transition 
at $T > T_{\rm RW} \approx 1.13T_{c0}$ and $\hat{\mu}^2=-(\pi/2)^2$.  
The value of $T_{\rm RW}$ is 178~MeV in the PNJL model, but $163\sim 166$MeV 
in LQC$_2$D data.  
In the present model, it is quite difficult to reproduce 
LQC$_2$D values of $T_{c0}$ and $T_{\rm RW}$ simultaneously. 
A fine tuning of the Polyakov potential ${\cal U}(\Phi)$ may be necessary.

\subsection{\rm Quark number density}

\begin{figure}[htb]
\begin{center}
\hspace{10pt}
\includegraphics[angle=0,width=0.445\textwidth]{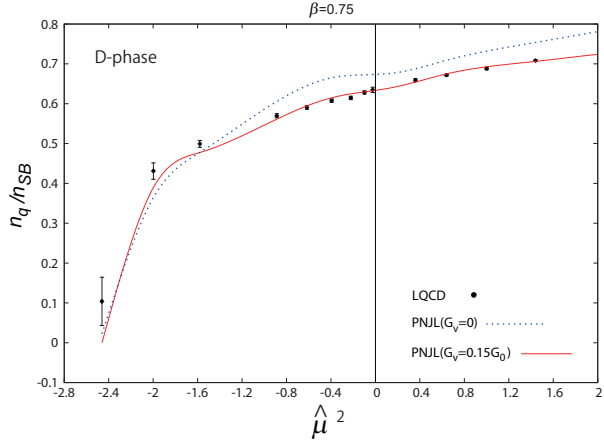}
\end{center}
\caption{$\hat{\mu}^2$-dependence of $n_q/n_{\rm SB}$ 
for $\beta=0.75~(T/T_{c0}=1.18)$.  
The dots with error bars are the results of LQC$_2$D data, 
while the solid (dotted) line corresponds to the result of PNJL model 
with $G_{\rm v}/G_0=0.15~(0)$.   
Note that $n_q/n_{\rm SB}$ at $\hat{\mu} =0$ is defined 
by $\displaystyle{\lim_{\hat{\mu}\to 0} n_q/n_{\rm SB}}$.}
\label{compare_n_beta_0.75}
\end{figure}

Figure~\ref{compare_n_beta_0.75} shows $\hat{\mu}^2$ dependence of 
$n_q/n_{\rm SB}$ at $\beta =0.75~(T/T_{c0}=1.18)$. 
As mentioned in the previous subsection, we have used LQC$_2$D 
on $n_q/n_{\rm SB}$ at $\beta =0.75$ and imaginary $\mu$ to determine 
the parameter set of the PNJL model. 
The parameter set thus determined well reproduces LQC$_2$D data even 
at real $\mu$. 
This ensures the assumption that the PNJL model is valid at high $T$. 
Also note that the PNJL model with $G_{\rm v}=0$ fails to reproduce LQC$_2$D 
data at real $\mu$, while the PNJL model with $G_{\rm v}$ agrees with 
the LQC$_2$D data even at real $\mu$.  
The imaginary chemical potential matching approach is thus a promising method. 

\begin{figure}[htb]
\begin{center}
\hspace{10pt}
\includegraphics[angle=0,width=0.445\textwidth]{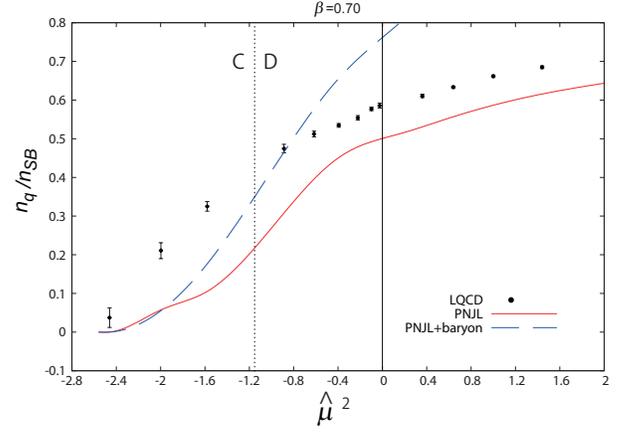}
\end{center}
\caption{$\hat{\mu}^2$-dependence of $n_q/n_{\rm SB}$ for $\beta=0.70~(T/T_{c0}=1.06)$. 
The dots with error bars are the results of LQC$_2$D data, while 
the solid (dashed) line corresponds to the result of the PNJL model 
(the PNJL+baryon model).  
On the left (right) side of the thin dotted line, 
the system is in the C-phase (D-phase). 
}
\label{compare_n_beta_0.70}
\end{figure}

Figure~\ref{compare_n_beta_0.70} shows the same as Fig.~\ref{compare_n_beta_0.75} but for $\beta =0.70~(T/T_{c0}=1.06)$. 
On the left (right) side of the vertical thin dotted line, 
the system is in the C-phase (D-phase). 
The PNJL result (solid line) underestimates LQC$_2$D data sizably 
in the C-phase. To improve this, we consider free ``baryons" and assume that 
the baryons have the same degrees of freedom as PS mesons, 
and add a contribution of free baryon gas to the PNJL results. 
Note that, in QC$_2$D, the baryon is a boson consisting of two quarks. 
According to Ref.~\cite{Muroya_nc2}, the scalar baryon with the same degree 
of freedom as the PS meson has almost the same mass as the PS meson.  
Hence, we use the baryon mass $m_{\rm B}=m_{\rm ps}=616$MeV. 
In this way, the baryon contribution to $n_q$ is given by
\begin{eqnarray}
n_{q,\rm B}=2g\int {d^3p\over{(2\pi )^3}}\left[{1\over{e^{\beta (E_{\rm B}-2\mu)}-1}}
-{1\over{e^{\beta (E_{\rm B}+2\mu)}-1}}\right], 
\nonumber\\
\label{nq_baryon}
\end{eqnarray}
where $E_{\rm B}=\sqrt{p^2+{m_{\rm B}}^2}$, $g=3$ is the degree of freedom and the factor 2 in front of $g$ comes from the fact that the baryon is composed of two quarks. 
This modification improves agreement with LQC$_2$D particularly in the C-phase, but not in the D-phase. This implies that baryons disappear in the D-phase 
at least partially. 

\begin{figure}[htb]
\begin{center}
\hspace{10pt}
\includegraphics[angle=0,width=0.445\textwidth]{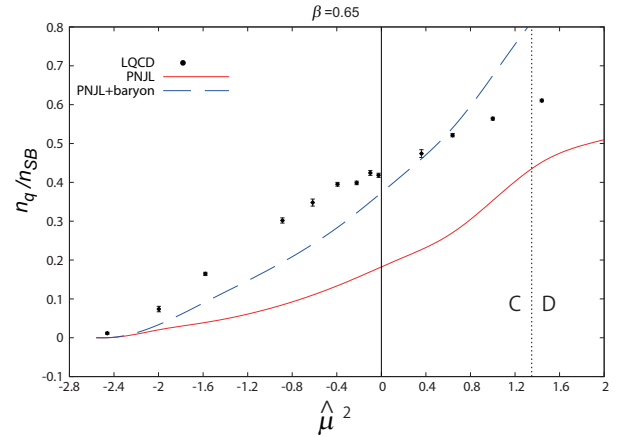}
\end{center}
\caption{$\hat{\mu}^2$-dependence of $n_q/n_{\rm SB}$ for 
$\beta=0.65~(T/T_{c0}=0.96)$. 
For the definition of lines, see the caption of 
Fig.~\ref{compare_n_beta_0.70}. 
 }
\label{compare_n_beta_0.65}
\end{figure}

Figure~\ref{compare_n_beta_0.65} shows the same as Fig.~\ref{compare_n_beta_0.75} but for $\beta =0.65~(T/T_{c0}=0.96)$. 
The system is in the C-phase on the left side of the thin dotted line, 
but in the D-phase on the right side. 
Again, the PNJL model undershoots LQC$_2$D in the C-phase, but 
the PNJL+baryon model almost reproduces the LQC$_2$D data in the C-phase, 
although the latter model overshoots LQC$_2$D data in the D-phase. 
Thus, the baryon may disappear in the D-phase. 

More precisely, the PNJL+baryon model overestimates LQC$_2$D data also in 
the C-phase near the thin dotted line. 
This fact may imply that a repulsive force works between baryons 
there. It is well known that such a repulsive force 
suppresses the baryon number density in realistic nuclear matter.

\begin{figure}[htb]
\begin{center}
\hspace{10pt}
\includegraphics[angle=0,width=0.445\textwidth]{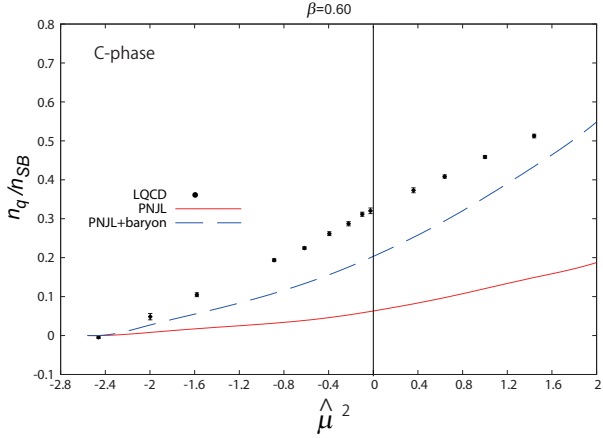}
\end{center}
\caption{$\hat{\mu}^2$-dependence of $n_q/n_{\rm SB}$ for 
$\beta=0.60~(T/T_{c0}=0.88)$. 
For the definition of lines, see the caption of 
Fig.~\ref{compare_n_beta_0.70}. 
}
\label{compare_n_beta_0.60}
\end{figure}

Figure~\ref{compare_n_beta_0.60} shows the same as Fig.~\ref{compare_n_beta_0.75} but for $\beta =0.60~(T/T_{c0}=0.88)$. 
The system is in the C-phase and the PNJL model largely underestimates 
LQC$_2$D data, but it is improved by the PNJL+baryon model. 
We can therefore conclude that baryon effects play an important contribution 
to $n_q$, whenever the system is in the C-phase.

\subsection{\rm Chiral condensate}

As mentioned in Sec. \ref{LSP}, the renormalization is necessary 
for the chiral condensate. 
The PNJL results are then simply multiplied by a normalization factor 1.92 
so that the results can reproduce LQC$_2$D data at $\beta =0.75$ 
($T/T_{c0}=1.18$) and $\hat{\mu}^2=\left(i{\pi/2}\right)^2$. 
This choice of normalization is natural, since the PNJL model is 
a quark model and is expected to be more reliable at higher $T$.

\begin{figure}[htb]
\begin{center}
\hspace{10pt}
\includegraphics[angle=0,width=0.445\textwidth]{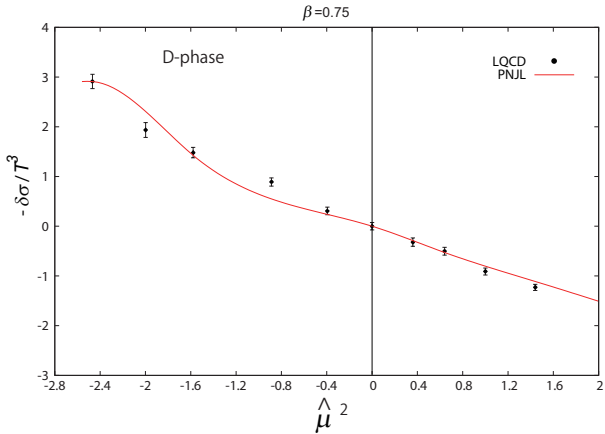}
\end{center}
\caption{$\hat{\mu}^2$-dependence of $\delta{\sigma}/T^3$ at $\beta=0.75~(T/T_{c0}=1.18)$. 
The dots with error bars are the results of LQC$_2$D data, while 
the solid line corresponds to the result of PNJL model. 
The PNJL result is multiplied by a normalization factor 1.92. }
\label{compare_chiral_beta_0.75}
\end{figure}

Figure~\ref{compare_chiral_beta_0.75} shows $\hat{\mu}^2$ dependence 
of $\delta \sigma$ at $\beta =0.75~(T/T_{c0}=1.18)$. 
At this temperature, the system is in the D-phase, and 
the PNJL model well reproduces LQC$_2$D data even at real $\mu$.

\begin{figure}[htb]
\begin{center}
\hspace{10pt}
\includegraphics[angle=0,width=0.445\textwidth]{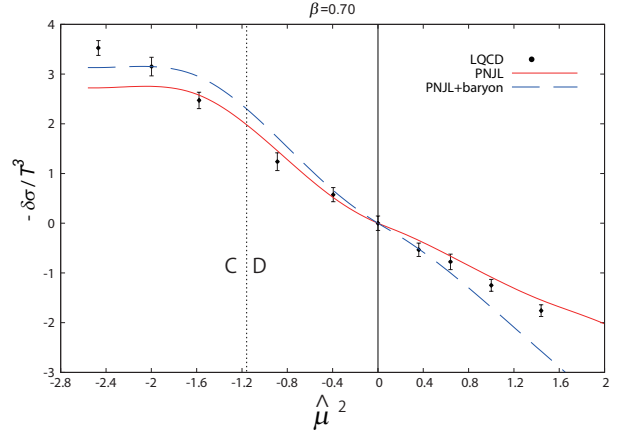}
\end{center}
\caption{$\hat{\mu}^2$-dependence of $\delta{\sigma}$ at $\beta=0.70~(T/T_{c0}=1.06)$. 
The dots with error bars are the results of LQC$_2$D data, while 
the solid (dashed) line corresponds to 
the results of the PNJL (PNJL+baryon) model.  
The model results are multiplied by a normalization factor 1.92. 
The system is in the C-phase on the left side of the thin vertical 
dotted line, but in the D-phase on the right side. 
}
\label{compare_chiral_beta_0.70}
\end{figure}

\begin{figure}[htb]
\begin{center}
\hspace{10pt}
\includegraphics[angle=0,width=0.445\textwidth]{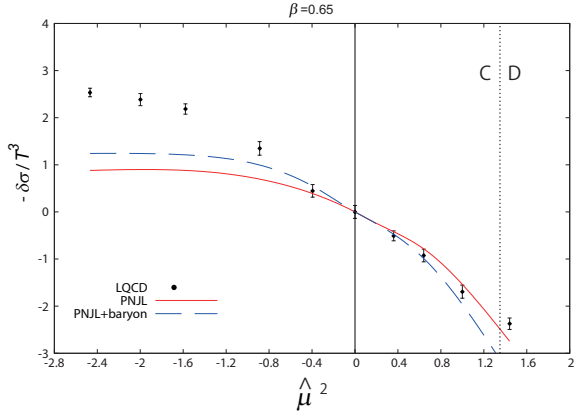}
\end{center}
\caption{$\hat{\mu}^2$-dependence of $\delta{\sigma}$ at $\beta=0.65~(T/T_{c0}=0.96)$. 
For the definition of lines, see the caption 
of Fig.~\ref{compare_chiral_beta_0.70}. 
}
\label{compare_chiral_beta_0.65}
\end{figure}

Figures~\ref{compare_chiral_beta_0.70} and \ref{compare_chiral_beta_0.65} 
show $\hat{\mu}^2$-dependence of $\delta \sigma$ at $\beta =0.70~(T/T_{c0}=1.06)$ and $\beta=0.65~(T/T_{c0}=0.96)$, respectively. 
On the left side of the thin vertical dotted line the system is in the C-phase, while it is in the D-phase on the right side. 
The PNJL model (solid line) is consistent with  LQC$_2$D data in the D-phase, 
but not in the C-phase. In order to improve this disagreement in the C-phase, 
we add baryon effects to the PNJL model again: 
\begin{eqnarray}
\sigma_{\rm B}&=&{\partial m_{\rm B}\over{\partial m}}g\int {d^3p\over{(2\pi )^3}}{m_{\rm B}\over{E_{\rm B}}}\left[{1\over{e^{\beta (E_{\rm B}-2\mu)}-1}}\right.
\nonumber\\
&&\left.+{1\over{e^{\beta (E_{\rm B}+2\mu)}-1}}\right], 
\label{sigma_baryon}
\end{eqnarray}
where we assume 
\bea
{\partial m_{\rm B}\over{\partial m}}=2 ,
\eea
since a naive constituent quark model gives this value. 
As shown in Figs~\ref{compare_chiral_beta_0.70} 
and \ref{compare_chiral_beta_0.65}, 
the PNJL+baryon model (dashed line) is more consistent with LQC$_2$D data than 
the PNJL model (solid line) in the C-phase, but less consistent 
in the D-phase. This means that baryon effects are significant only 
in the $C$-phase.

\begin{figure}[htb]
\begin{center}
\hspace{10pt}
\includegraphics[angle=0,width=0.445\textwidth]{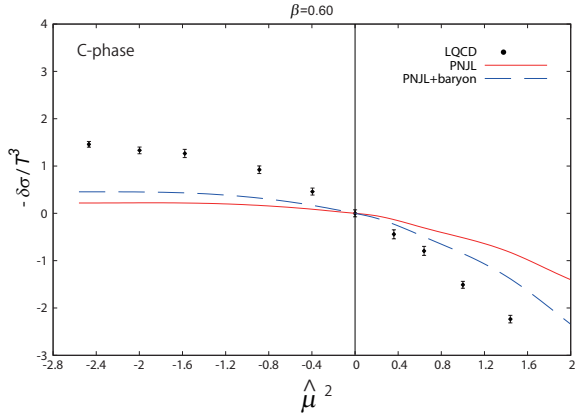}
\end{center}
\caption{$\hat{\mu}^2$-dependence of $\delta{\sigma}/T^3$ at $\beta=0.60~(T/T_{c0}=0.88)$. 
For the definition of lines, see the caption 
of Fig.~\ref{compare_chiral_beta_0.70}. 
}
\label{compare_chiral_beta_0.60}
\end{figure}

Figure~\ref{compare_chiral_beta_0.60} shows $\hat{\mu}^2$-dependence 
of $\delta \sigma$ at $\beta =0.60$ $(T/T_{c0}=0.88)$. 
At this temperature, the system is in the C-phase. 
The PNJL+baryon model (dashed line) yields better agreement with LQC$_2$D 
than the PNJL model (solid line) in the whole region 
of $\hat{\mu}^2$; note that $\delta \sigma$ is always zero at $\hat{\mu}=0$ 
by the definition (\ref{E_delta_sigma}). 
Thus, baryon effects are important in the C-phase 
not only for $n_q/n_{\rm SB}$ 
but also for $\delta \sigma$.

\subsection{\rm Polyakov loop}

\begin{figure}[htb]
\begin{center}
\hspace{10pt}
\includegraphics[angle=0,width=0.445\textwidth]{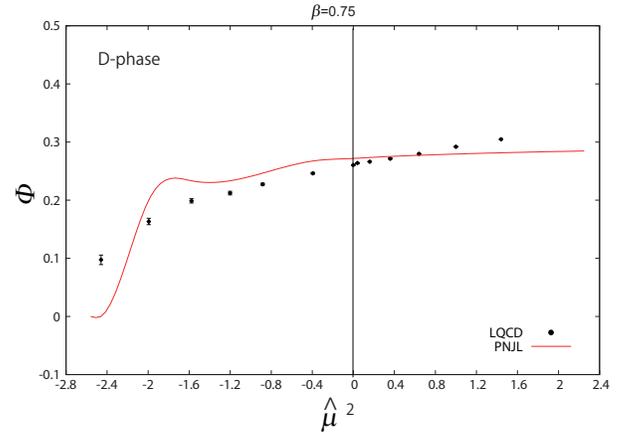}
\end{center}
\caption{$\hat{\mu}^2$-dependence of $\Phi$ at $\beta=0.75~(T/T_{c0}=1.18)$.   
The dots with error bars are the results of the LQC$_2$D data, while 
the solid line corresponds to the result of PNJL model. 
The PNJL result is multiplied by the normalization factor 0.304. }
\label{compare_pol_beta_0.75}
\end{figure}

Figure~\ref{compare_pol_beta_0.75} shows $\hat{\mu}^2$ dependence of $\Phi$ 
at $\beta =0.75$.  
As mentioned in Sec. \ref{Sec:comparison}, the PNJL result is multiplied 
by a normalization factor 0.304 for $\Phi$.  
At this temperature, the system is in the D-phase. 
The PNJL model (solid line) well reproduces LQC$_2$D data except for 
the vicinity of the first-order RW transition.

\begin{figure}[htb]
\begin{center}
\hspace{10pt}
\includegraphics[angle=0,width=0.445\textwidth]{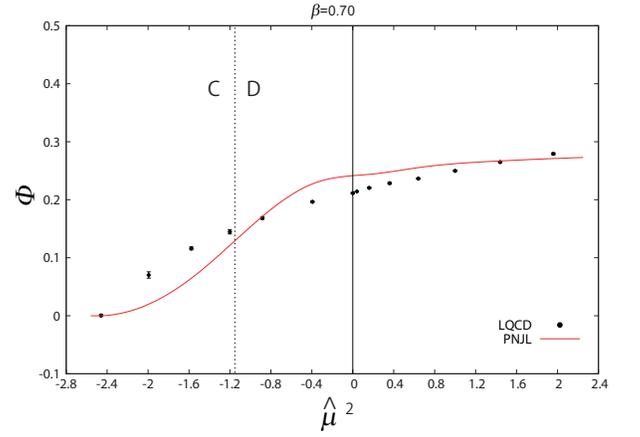}
\end{center}
\caption{$\hat{\mu}^2$-dependence of $\Phi$ at $\beta=0.70~(T/T_{c0}=1.06)$.   
The dots with error bars are the results of the LQC$_2$D data, while the solid line is the result of PNJL model. 
The PNJL result is multiplied by the normalization factor 0.304. 
The system is in the C-phase on the left side of the thin vertical dotted line, while it is in the D-phase on the right side.  
}
\label{compare_pol_beta_0.70}
\end{figure}

\begin{figure}[htb]
\begin{center}
\hspace{10pt}
\includegraphics[angle=0,width=0.445\textwidth]{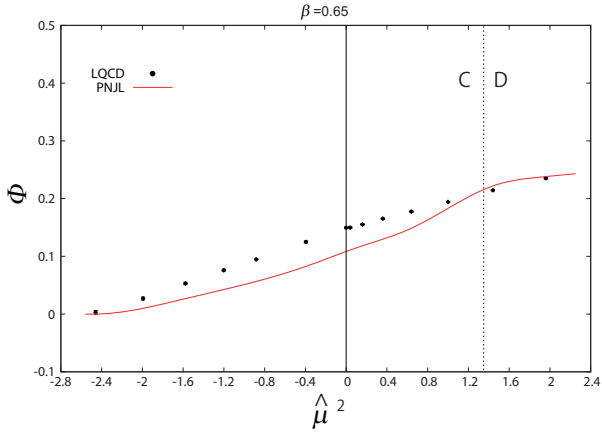}
\end{center}
\caption{$\hat{\mu}^2$-dependence of $\Phi$ at $\beta=0.65~(T/T_{c0}=0.96)$.  
For the definition of lines, see the caption of Fig.~\ref{compare_pol_beta_0.70}.  
}
\label{compare_pol_beta_0.65}
\end{figure}

Figures~\ref{compare_pol_beta_0.70} and~\ref{compare_pol_beta_0.65} show 
$\hat{\mu}^2$ dependence of $\Phi$ at $\beta =0.70$ and $\beta =0.65$, 
respectively.  
The PNJL model deviates from LQC$_2$D data to some extent in the C-phase except for the point at $\hat{\mu}^2=\left(i{\pi\over{2}}\right)^2$ where $Z_2$ symmetry restricts $\Phi$ to zero.


\begin{figure}[htb]
\begin{center}
\hspace{10pt}
\includegraphics[angle=0,width=0.445\textwidth]{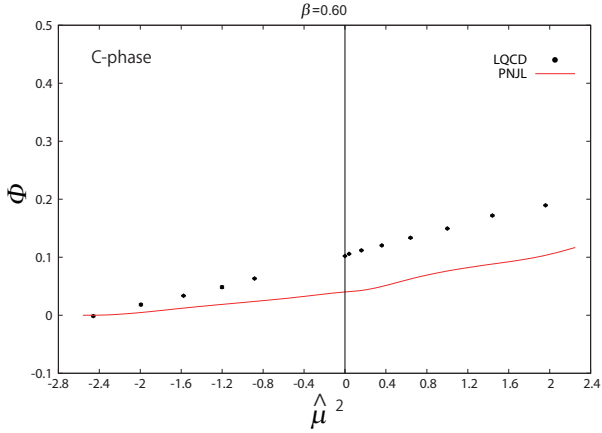}
\end{center}
\caption{$\hat{\mu}^2$-dependence of $\Phi$ at $\beta=0.60~(T/T_{c0}=0.88)$.  
For the definition of lines, see the caption of Fig.~\ref{compare_pol_beta_0.75}. 
}
\label{compare_pol_beta_0.60}
\end{figure}

Figures~\ref{compare_pol_beta_0.60} shows $\hat{\mu}^2$-dependence of $\Phi$ at $\beta =0.60$. 
At this temperature, the system is in the C-phase.  
The PNJL model deviates from LQC$_2$D data to some extent in the whole region 
except for the point at $\hat{\mu}^2=\left(i{\pi/{2}}\right)^2$ where $Z_2$ symmetry restricts $\Phi$ to zero.

Throughout all the analyses for $\Phi$, we can say that 
the PNJL model cannot reproduce LQC$_2$D data properly 
in the vicinity of RW transition and at lower $T$. 
A fine tuning of ${\cal U}(\Phi)$ may be necessary to 
solve this problem.

\subsection{\rm Phase diagram}

\begin{figure}[htb]
\begin{center}
\hspace{10pt}
\includegraphics[angle=0,width=0.445\textwidth]{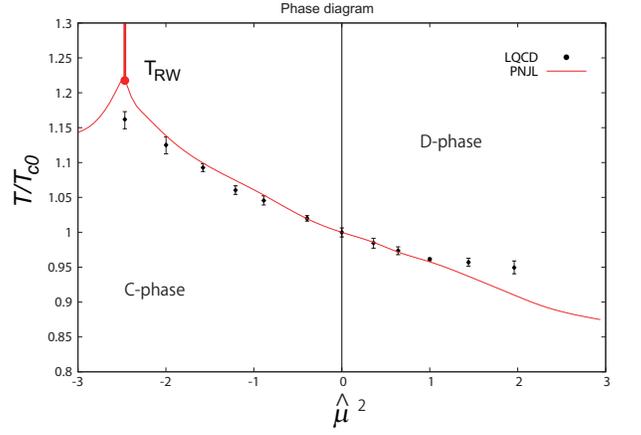}
\end{center}
\caption{Phase diagram in $\hat{\mu}^2$--$T$ plane.  
The dots with error bars represent the Pseudo-critical line of LQC$_2$D,  
while the solid line corresponds to the phase diagram of PNJL model. 
The vertical thick solid line is the RW transition line determined with 
the PNJL model. 
}
\label{compare_phase}
\end{figure}

Figure~\ref{compare_phase} shows the phase diagram in $\hat{\mu}^2$--$T$ plane. The PNJL model (solid line) reproduces LQC$_2$D data well in a range of 
$\hat{\mu}^2=-2\sim 1$, but it overshoots LQC$_2$D data to some extent 
near the RW transition line and undershoots LQC$_2$D data in the large $\hat{\mu}^2$ region of 
$\hat{\mu}^2 \approx (1.2)^2$.   
It is an interesting question whether the PNJL model can reproduce 
LQC$_2$D data in the large $\hat{\mu}^2$ region as soon as 
the model is improved  
to reproduce LQC$_2$D data near the RW transition line.  

\section{Summary}
\label{Summary}

We studied the phase structure of QC$_2$D at both real and 
imaginary $\mu$ by using an $8^3 \times 4$ lattice with the renormalization-group improved Iwasaki gauge action~\cite{Iwasaki,Iwasaki3} and the clover-improved  two-flavor Wilson fermion action~\cite{clover-Wilson}.  
The Polyakov loop, the chiral condensate and the quark number density were 
calculated at $0.86 \le T/T_{c0} \le 1.18$ and 
$-(\pi/2)^2 \le \hat{\mu}^2 \le (1.2)^2$. 
These quantities are smooth at $\hat{\mu}=0$, as expected. 
This guarantees that the analytic continuation 
of physical quantities from imaginary $\hat{\mu}$ to real $\hat{\mu}$ 
is possible. 

Accuracy of the analytic continuation was tested 
in Refs. \cite{Cea_1,Cea_2} with staggered fermions. 
In this paper we have made similar analyses 
with clover-improved Wilson fermions 
by assuming a polynomial series 
of $\hat{\mu}$ in the deconfinement phase and a Fourier series 
in the confinement phase, where 
coefficients of the series were determined at imaginary $\mu$. 
As for the quark number density at $T/T_{c0}=1.18$ corresponding to 
the deconfinement phase, the polynomial series up to $\hat{\mu}^3$ 
well reproduces LQC$_2$D results in a wide range of 
$0 \le \hat{\mu}^2 \leq (1.2)^2$. 
At $T/T_{c0}=0.88$ corresponding to the confinement phase, 
the results of the lowest-order Fourier series $\sin(2\theta)$ are 
consistent with LQC$_2$D results 
in a range of $0 \le \hat{\mu}^2 < 0.8$. 
At $T/T_{c0}=0.96$ near the deconfinement transition, it is good only 
in $0 \le \hat{\mu}^2 < 0.4$. 
The analytic continuation is thus useful in the deconfinement 
and confinement regions, but less accurate in the transition region near 
$T/T_{c0}=1$ where the deconfinement crossover takes place somewhere 
in a range of $-(\pi/2)^2 \le \hat{\mu}^2 \le (1.2)^2$ as 
$\hat{\mu}^2$ increases with $T$ fixed. 
This is true for other quantities such as the Polyakov loop 
and the chiral condensate. 

We have tested the validity of the PNJL model by comparing model results 
with LQC$_2$D ones, where the model parameters are fitted 
to the quark number density at $T/T_{c0}=1.18$ and imaginary $\mu$. 
As for the transition line of deconfienment crossover, 
the model result agrees with LQC$_2$D one. More precisely, 
the agreement is not perfect in 
the vicinity of RW transition line and the large-$\hat{\mu}^2$ region of 
$({\hat \mu})^2 \approx (1.2)^2$. It is interesting 
whether the PNJL model can reproduce LQC$_2$D data 
in the large $\hat{\mu}^2$ region as soon as the model is improved  
to reproduce LQC$_2$D data near the RW transition line. 
A possible candidate of the improvement is a fine tuning of 
the Polyakov-loop potential ${\cal{U}}(\Phi)$.

In the deconfinement region of $T/T_{c0}=1.18$, the PNJL model yields good 
agreement with LQC$_2$D data at both real and imaginary $\mu$ 
for the quark number density, 
the chiral condensate and the Polyakov loop. 
The agreement particularly at real $\mu$ indicates 
that the PNJL model is reliable in the deconfinement region. 
In the transition region of $T/T_{c0} \approx 1$, 
the agreement of the PNJL model with LQC$_2$D data is not perfect. 
As for the quark number density and the chiral condensate, 
however, the deviation can be reduced 
in the confinement area appearing at smaller $\hat{\mu}^2$ 
by introducing baryon degree of freedom to the PNJL model. 
In the deconfinement area appearing at larger 
$\hat{\mu}^2$, on the contrary, the model overestimates 
LQC$_2$D results if the baryon contribution is taken into account. 
This means that baryons disappear at least partially 
in the deconfinement area. 
Also in the confinement region of $T/T_{c0}=0.88$, 
the baryon degree of freedom is important. 
As for the Polyakov loop, 
the disagreement between PNJL and LQC$_2$D results in the confinement area 
cannot be solved by the baryon contribution.   
Of course, this comes from the fact that the Polyakov-loop potential is 
not changed by the baryon contribution in the present framework. 
The improvement of PNJL model along this line is interesting.

The present analysis also shows that the vector-type four-quark interaction 
is necessary to explain LQC$_2$D data on the quark number density. 
This fact indicates that also in the realistic case of three colors 
the strength of vector-type interaction can be determined from LQCD data 
at imaginary $\mu$~\cite{Sakai_vector,Sakai_para,Kashiwa_nonlocal}.  
Although the lattice we used is quite coarse and the number of LQC$_2$D data is limited, the present results surely show that the imaginary-$\mu$ 
matching approach~\cite{Kashiwa_meson} is a promising approach 
to thermodynamics of QCD at finite $\mu$.

In this paper, we consider only the $\mu$-region where the diquark condensate 
is zero. Quantitative check of our effective model 
in the region~\cite{Hands,Hands_2,Cotter} is also interesting.

\noindent
\begin{acknowledgments} 
T. M. and H. K. thank H. Yoneyama, A. Sugiyama and M. Tachibana for valuable comments. 
T. S., J.T., K.K., H.K., M.Y and A.N. are supported   
by JSPS KAKENHI No. 23749194, No. 25-3944, No. 26-1717, No. 26400279, No. 26400278, and Nos. 24340054 and 26610072, respectively. 
The numerical calculations were performed by using the NEC SX-9 and SX-8R at CMC, Osaka University, and by using the RIKEN Integrated Clusters (RICC) facility. 
\end{acknowledgments}
\twocolumngrid

\appendix
\section{Coefficients of fitting functions for analytic continuation}
\label{Parameters of fitting functions}

We present the coefficients of analytic functions determined 
from LQC$_2$D data at imaginary $\mu$ and the $\chi^2/{\rm d.o.f}$ 
for the fitting.

\begin{table}[htb]
\begin{center}
\scalebox{0.7}[0.7]{
\begin{tabular}{cccccc}
\hline\hline
(a) & $\beta =0.75$ & & & & \\
\hline
observable & function & $A$ & $B$ & $C$ & $\chi^2/{\rm d.o.f}$\\
\hline
${\rm Im}(n_q)$  &{\rm Eq.}(\ref{function_density_high})& 3.61014(1466) &  0.84703(2265) & &2.914  \\
${\rm Im}(n_q)$  &{\rm Eq.}(\ref{function_density_high_5})& 3.57904(1989) &  0.70990(6347)& -0.08459(3657) &  2.566   \\
$\delta\sigma$ &{\rm Eq.}(\ref{function_pol_chiral_high}) &-0.07116(5561) &  -1.07870(5010) &  &  2.337 \\
$\delta\sigma$ &{\rm Eq.}(\ref{function_pol_chiral_high_4}) & 0.01304(8702)&  -0.7336(0190)& 0.1571(8106)&  1.448 \\
$\Phi$ &{\rm Eq.}(\ref{function_pol_chiral_high})& 0.26153(70) &  0.04379(12) &  &  13.25 \\
$\Phi$ &{\rm Eq.}(\ref{function_pol_chiral_high_4})& 0.25994(73) &  0.02410(310)& -0.01331(191) &  4.534 \\
\hline\hline
(b) & $\beta =0.70$  & & & & \\
\hline
observable & function & $A$ & $B$ & $C$ & $\chi^2/{\rm d.o.f}$\\
\hline
${\rm Im}(n_q)$  &{\rm Eq.}(\ref{function_density_high})& 3.32812(2089) &  1.07313(5618) &  &  1.077  \\
${\rm Im}(n_q)$  &{\rm Eq.}(\ref{function_density_high_5})& 3.34986(2818) &  1.2644(1755) & 0.24168(21017) &  1.016 \\
$\delta\sigma$ &{\rm Eq.}(\ref{function_pol_chiral_high}) & 0.00868(13029) &  -1.39546(25909) &  & 0.018  \\
$\Phi$ &{\rm Eq.}(\ref{function_pol_chiral_high_4})& 0.21241(112) &  0.04621(279)&  & 5.590  \\
\hline\hline
(c) & $\beta =0.65$  & & & & \\
\hline
observable & function & $A$ & $B$ & $C$ & $\chi^2/{\rm d.o.f}$\\
\hline
${\rm Im}(n_q)$  &{\rm Eq.}(\ref{function_density_low}) & 1.31566(759) &  &  & 23.80   \\
${\rm Im}(n_q)$  &{\rm Eq.}(\ref{function_density_low_4})& 1.44376(1216) & -0.11951(886) & &4.075\\
$\delta\sigma$ &{\rm Eq.}(\ref{function_chiral_low})  &1.1107(5784)& -1.33556(6933) &  & 2.155   \\
$\delta\sigma$ &{\rm Eq.}(\ref{function_chiral_low_4}) &1.05958(6059)&  -1.27686(7235) & 0.19765(6969) & 0.1933\\
$\Phi$ &{\rm Eq.}(\ref{function_pol_low})& 0.15644(85) &  &  &7.920 \\
$\Phi$ &{\rm Eq.}(\ref{function_pol_low_3})& 0.15375(94) & -0.00461(68) & &0.442 \\
\hline\hline
(d) & $\beta =0.60$ & & & & \\
\hline
observable & function & $A$ & $B$ & $C$ & $\chi^2/{\rm d.o.f}$\\
\hline
${\rm Im}(n_q)$  &{\rm Eq.}(\ref{function_density_low}) &0.93599(80)  &  &  &1.189   \\
${\rm Im}(n_q)$  &{\rm Eq.}(\ref{function_density_low_4}) & 0.94541(838) & -0.01438(74) &  & 0.877\\
$\delta\sigma$ &{\rm Eq.}(\ref{function_chiral_low})  & 0.69931(3257) & -0.71395(3973) &   & 0.2853    \\
$\delta \sigma$ &{\rm Eq.}(\ref{function_chiral_low_4}) & 0.69633(3414) & -0.71059(4087) & 0.01373(3913) &  0.3394 \\
$\Phi$ &{\rm Eq.}(\ref{function_pol_low})& 0.10258(38) &  &  &  3.826 \\
$\Phi$ &{\rm Eq.}(\ref{function_pol_low_3})& 0.10404(52) & -0.00195(46) &  &0.4438   \\
\hline\hline
\end{tabular}}
\end{center}
\caption{Coefficients of analytic functions and $\chi^2/{\rm d.o.f}$ 
for quark number density, chiral condensate and the Polyakov loop 
at (a) $\beta =0.75$, (b) $\beta =0.70$, (c) $\beta =0.65$, 
and (d) $\beta=0.60$. 
The coefficients are determined from LQC$_2$D data at imaginary $\mu$. 
The fitting range is $\hat{\mu}^2 =-1.15\sim 0$ for (b) and 
$\hat{\mu}^2 =-(\pi /2)^2\sim 0$ for the other cases. }
\label{para_quantity}
\end{table}

\begin{table}[htb]
\begin{center}
\scalebox{0.8}[0.8]{
\begin{tabular}{ccccc}
\hline\hline
function & $A$ & $B$ & $C$ & $\chi^2/{\rm d.o.f}$\\
\hline
{\rm Eq.}(\ref{function_pol_chiral_high})& 0.66802(204) & -0.02868(167) &      &     0.109\\
{\rm Eq.}(\ref{function_pol_chiral_high})& 0.66871(257) & -0.02637(557) &    0.00113(260) & 0.089\\
\hline\hline
\end{tabular}
}
\end{center}
\caption{Coefficients of analytic functions and $\chi^2/{\rm d.o.f}$ 
for the pseudo-critical line.   
The coefficients are determined from LQC$_2$D data at imaginary $\mu$. 
The fitting range is $\hat{\mu}^2 =-(\pi /2)^2\sim 0$ for all the cases. }
\label{para_phase}
\end{table}

~

\clearpage


\end{document}